\documentclass[reprint,superscriptaddress,amsmath,amssymb,aps,pra]{revtex4-1}
\pdfoutput=1

\usepackage{color}

\usepackage{graphicx}
\usepackage{dcolumn}
\usepackage{bm}
\usepackage{braket}
\usepackage{hyperref}

\begin{document}

\title{Real-time calibration with spectator qubits}

\author{Swarnadeep Majumder}
\affiliation{Departments of Electrical and Computer Engineering, Chemistry, and Physics, Duke University, Durham, NC 27708 USA}%
\author{Leonardo Andreta de Castro}\email{leoadec@leoad.ec}
\affiliation{Departments of Electrical and Computer Engineering, Chemistry, and Physics, Duke University, Durham, NC 27708 USA}%
\affiliation{Present address: Q-CTRL Pty Ltd, Sydney, NSW Australia}
\author{Kenneth R. Brown}
\affiliation{Departments of Electrical and Computer Engineering, Chemistry, and Physics, Duke University, Durham, NC 27708 USA}%

\date{9 February 2020}

\begin{abstract}
Accurate control of quantum systems  requires precise measurement of the parameters that govern the dynamics, including control fields and interactions with the environment.  Parameters will drift in time and experiments interleave protocols that perform parameter estimation with protocols that measure the dynamics of interest.  Here we specialize to a system made of qubits where the dynamics correspond to a quantum computation. We  propose setting aside some qubits, which we call spectator qubits, to be measured periodically during the computation, to act as probes of the changing experimental and environmental parameters. By using control strategies that minimize the sensitivity of the qubits involved in the computation,  we can acquire sufficient information from the spectator qubits to update our estimates of the parameters and improve our control. As a result, we can increase the length of experiment where the dynamics of the data qubits are highly reliable. In  particular, we simulate how spectator qubits can keep the error level of operations on data qubits below a $10^{-4}$ threshold in two scenarios involving coherent errors: a classical magnetic field gradient dynamically decoupled with sequences of two or four $\pi$-pulses, and laser beam instability detected via crosstalk with neighboring atoms in an ion trap.
\end{abstract}

\maketitle

\section{Introduction}
One of the key challenges in constructing a quantum computer is keeping the error rate under an acceptable threshold, which will be a requirement even for future fault--tolerant quantum computation~\cite{AharonovSTOC1997,KitaevRMS1997,KnillScience1998,GottesmanPRA1998,KnillNature2005,AliferisPRL2007,TerhalRMP2015}.
The optimal control strategy for each quantum gate depends on the parameters that characterize the underlying error channel $\mathcal{E}$.
There has been an increasing interest in tailoring control strategies to the error channel, such as variability-aware qubit allocation and movement~\cite{TannuASPLOS2019}, optimal quantum control using randomized benchmarking~\cite{KellyPRL2014}, robust phase estimation~\cite{KimmelPRA2015}, noise-adaptive compilation~\cite{MuraliASPLOS2019}, and quantum error-correcting codes designed for biased noise~\cite{NappQIC2013,TuckettPRL2018,TuckettPRX2019,MuyuanPRX2019}.

Although an initial calibration may be sufficient for simpler devices, a fully functional quantum computer will have to deal with the possibility of assessing changes in the error parameters in real time.
Many reduction techniques have been proposed for errors that vary slowly in time, such as composite pulses~\cite{LevittPNMRS1986,WimperisJMR1994,BrownPRA2004,TorosovPRA2011,VitanovPRA2011,BandoJPSJ2013,GenovPRL2014,LowPRA2014,KabytayevPRA2014,MerrillACP2014,LowPRX2016,CohenPRA2016}, optimal control~\cite{MontangeroPRL2007,GraceJMO2007,SchulteJPB2011,MachnesPRL2018}, dynamical decoupling~\cite{ViolaPRL1999,ViolaPRL2003,SouzaPRL2010,SouzaPTRSA2012,QuirozPRA2013,PokharelPRL2018}, and dynamically corrected gates~\cite{KhodjastehPRL2009,KhodjastehPRA2009,GreenPRL2012}.
In this work, we analyze the use of a subset of qubits -- called spectator qubits -- to perform real-time recalibration.

Spectator qubits probe directly the sources of error and thus do not need to interact with the data qubits, so they can be distinguished from
ancilla qubits used for syndrome extraction \cite{ShorPRA1995,TerhalRMP2015} in quantum error correction.
As long as the error channel of the spectator qubits is correlated to the error channel of the data qubits, it is possible to estimate $\mathcal{E}$ by measuring the spectators.
Although sensor networks~\cite{QianPRA2019}, machine leaning techniques~\cite{MavadiaNatureComms2017,GuptaPRApplied2018}, and even spectator qubits~\cite{GuptaARXIV2019} have been proposed to keep track of error parameters that vary in space or time, more often than not these techniques are not suitable for real-time calibration because of how long it takes to extract useful information about the error parameters from the experimental data. Here we describe the complete feedback loop between the information extracted from the spectator qubits and the recalibration of the control strategy on the data qubits, estimating how this information can positively impact the control protocol.
When the necessity for feedback is taken into account, acquisition of information via the spectator qubits has to be sufficiently fast such that the rate of errors in the data qubit does not exceed the rate at which the parameters are being estimated.
Such feedback schemes could in principle deal with general classes of errors, but in this work we will limit our discussion to particularly damaging coherent errors.

We illustrate the difficulty of using feedback against coherent errors with a simple example.
Consider constant overrotations around the $x$-axis characterized by the error parameter $\theta$.
If the error rate is the same as the rate of acquisition of information, the estimate of $\theta$ after $N$ overrotations will have an imprecision proportional to $N^{-1/2}$.
For this reason, any attempt to correct the error with the inverse unitary will result in an extant error that still grows with $O(N^{1/2})$:
\begin{equation}
    e^{i N \theta \sigma_x}
    e^{-i N \left[ \theta + O(N^{-1/2}) \right] \sigma_x}
    = e^{i O(N^{1/2}) \sigma_x}.
\end{equation}
This kind of difficulty is common to coherent errors in general, but can be contained with the help of quantum control techniques that reduce the speed with which the errors accumulate in the data qubits.
{  Other ways of balancing the increase of errors with a sufficient fast acquisition of information are: usage of different species of qubits for data and spectators, and application of different rates of measurements in order to reach the Heisenberg limit.
Here, we will focus on the first strategy of making the data qubits less sensitive to errors via control strategies, and leave the other methods to be explored in future work.
}

In this work, we propose that real-time calibration with spectator qubits can in principle improve the fidelity of any system undergoing coherent errors, as long as: (1) the information available to the spectator qubits is sufficient to keep track of the rate of change of the error parameters; and (2) we have a quantum control method capable of sufficiently suppressing the speed with which the coherent errors accumulate in the data.
{  The general setup that we will consider is the one illustrated in Fig. 1, where spectator qubits embedded in the same architecture as the data qubits are measured periodically to determine the error profile.
The errors might change in space and time, but as long as they have some correlation between neighboring qubits, we can extrapolate an error field that contains information about the error profile in the data qubits.
A classical apparatus then uses the information about this profile and about the results of the spectator measurements to decide the best control strategy. }
Here, we present the theoretical framework for studying multipartite systems composed of spectator qubits and data qubits in presence of coherent errors, propose some applications, and present analytical and numerical simulations of the performance of the spectator qubits.

\begin{figure}[htb]
\centering
\includegraphics[width=\columnwidth]{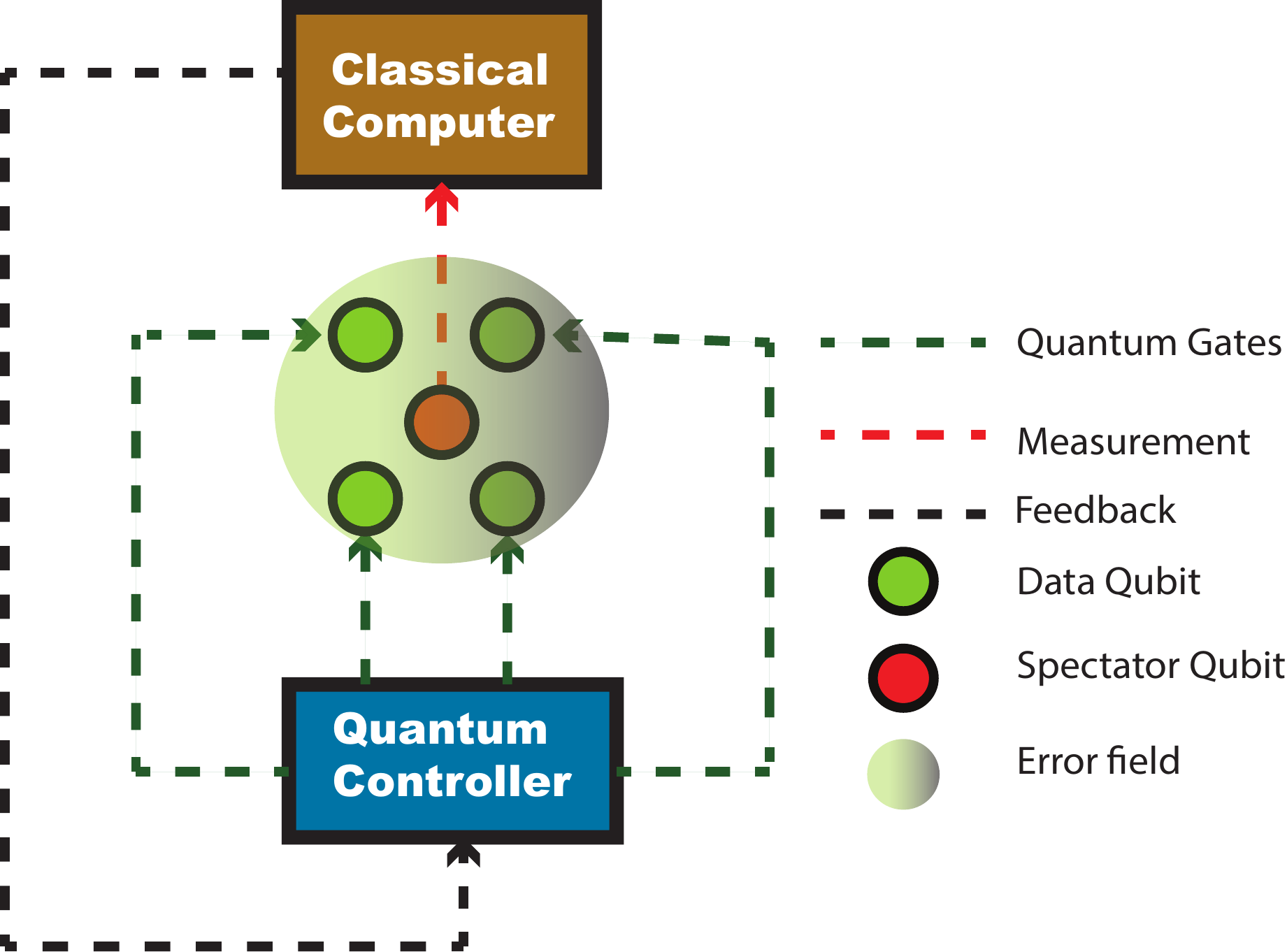}
\caption{
\textbf{General feedback loop setup for spectator qubits.} Information from the error field is acquired via measurements of the spectator qubits (red). This is analysed by a classical apparatus, which then updates the optimal control strategy for the data qubits (green).}
\end{figure}

\section{Results}

\subsection{Theoretical limits}

{

The purpose of spectator qubits is to obtain information about an error parameter while its value slowly drifts.
This problem can be approximated by a setting where the error parameter is assumed to be fixed within a series of time slots.
After the end of each time slot, the parameter changes according to a probability distribution.

In this work, we will assume that the drift is sufficiently slow that this error parameter does not change significantly during a measurement.
In this limit, the duration of the measurement process does not affect the result of the measurement, and we can assume it to be instantaneous.
However, the number $N$ of measurements that can be performed in the period of time where the error parameters remain fixed is still limited.

The precision with which we can learn about the drift from this series of $N$ measurements is limited by the Cram\' er--Rao bound~\cite{BraunsteinPRL1994}.
Calling our imprecision $\delta\vartheta$, the Cram\'er--Rao bound has the form:
\begin{equation}
    \left\langle | \delta\vartheta | \right\rangle \ge
    \frac{1}{\sqrt{N f_\theta}},
    \label{CramerRao}
\end{equation}
where $f_\theta$ is the Fisher information about the error parameter $\theta$ available to the system between each measurement.
The error parameter $\theta$ could in principle represent any kind of information about the error, but in the case of coherent errors it will often mean an overrotation angle.
Still in the context of coherent errors, if the system can be represented by a pure-state density matrix $\rho$, the Fisher information takes the specific form~\cite{BraunsteinPRL1994}:
\begin{equation}
    f_\theta = 4
    \mathrm{Tr} \left\{
        \rho \left( \frac{\partial \rho}{\partial \theta} \right)^2
    \right\}
    = \left( 1 - \langle \mathbf{\hat n} \cdot \boldsymbol\sigma \rangle^2 \right),
    \label{eq:Fisher}
\end{equation}
where we are expressing the unitary error operator in terms of the rotation axis $\mathbf{\hat n}$ as $U=e^{-i (\mathbf{\hat n} \cdot \boldsymbol\sigma) \theta/2}$, and the expectation value is taken for the initial state of the system.

While the imprecision decreases with $1/\sqrt{N}$, there are other resources that are able to improve our estimate faster than that.
Increasing the number of overrotations $L$ between each measurement reduces the imprecision by $1/L$ -- the so-called Heisenberg scaling -- a kind of precision that can also be achieved by using entangled qubits~\cite{WinelandPRA1992,BollingerPRA1996}.
These schemes increase the achievable precision by increasing the Fisher information between each measurement.
However, the scaling by the number of measurements will remain the same, $1/\sqrt{N}$.

Calling the error parameter during the $k$th time slot $\theta_k$, the optimal correction scheme against the coherent error $U(\theta_k)$ would be to simply apply its Hermitian conjugate, $U^\dagger(\theta_k)$.
Our actual strategy might not be as good as the ideal one, so this description will give us an upper limit for the possible recovery.
Moreover, if we only have access to an estimate $\hat \theta_k$ of $\theta_k$, our strategy will be limited by the amount of precision we can achieve in obtaining this estimate.
Although it is possible that our measurement scheme will not be able to saturate the Cram\' er--Rao bound, this will nevertheless provide an upper limit to how much precision we can achieve.
An example of a situation where the Cram\' er--Rao bound is not saturated occurs when we attempt to apply Robust Phase Estimation (RPE), a procedure that prescribes doubling the number of gates before each measurement~ \cite{RudingerPRL2017}, resulting in a precision that achieves the Heisenberg scaling~\cite{KimmelPRA2015} if there are no time constrains.
With time constraints, this scaling is not always true -- as discussed in Section V of the Supplementary Material.
Given this risk of underperformance, and the fact that robustness of RPE to time-dependent errors is not well-known~\cite{MeierPRA2019}, in this work we will only use a limited number of gate repetitions before measurement and concentrate our analysis on the scaling $1/\sqrt{N}$ that comes from varying the number of measurements $N$.

Regardless of how we obtain the estimate $\hat \theta_k$, once we have a reliable value for it, the best possible evolution after the optimal correction strategy is represented by the effective unitary $V(\phi)$:
\begin{equation}
    V (\phi) = U^\dagger(\hat \theta_k) U(\theta_k),
\end{equation}
where $\phi$ is an effective error parameter that depends essentially on the difference between $\theta_k$ and $\hat\theta_k$.
Using this, we can estimate the best possible process fidelity for a given $\phi$:
\begin{equation}
     F(\phi) = \left| \frac{\mathrm{Tr} \left\{ V^\dagger(\phi=0) V (\phi) \right\} }{\mathrm{Tr} \left\{ \mathbb{I} \right\}} \right|^2,
     \label{fid}
\end{equation}
which is proportional to the average fidelity \cite{PedersenPLA2007}.
Here, $\phi=0$ corresponds to perfect knowledge of the error parameter, allowing a perfect evolution of the system.
Additionally invoking the fact that the fidelity for a coherent error should be a continuous function of the angle, we can expand this fidelity as a power series around $F(\phi=0)$:
\begin{equation}
    F(\phi) = \sum_{n=0}^\infty \phi^n F^{(n)} (0),
\end{equation}
where we are using the following notation for the $n$th derivative of the fidelity:
\begin{equation}
    F^{(n)} (x) = \left. \frac{\mathrm{d}^n F}{\mathrm{d} \phi^n} \right|_{\phi =x}.
\end{equation}

By our choice of $\phi$, the point $\phi=0$ corresponds to perfect knowledge of the error, making the fidelity take its maximum value.
Therefore, $F(\phi=0)=1$, $F^{(1)} (\phi=0) =0$, and  $F^{(2)} (\phi=0) <0$.
Expanding the expression up to the second order and representing the extra terms as a Lagrange remainder, we find:
\begin{equation}
    F(\phi) =
    1 + \frac{1}{2} \phi^2 F^{(2)}(0) + \frac{1}{2} \int_0^\phi \mathrm{d} x \; (x-\phi)^2 F^{(3)} (x).
    \label{TaylorExpansion}
\end{equation}
Suppose $\phi_n$ is the effective error parameter if we do not update our estimates using the spectator qubits, whereas $\phi_s$ is the effective error parameter if we use spectator qubits to acquire information.
The feedback loop will be successful if the fidelity obtained using spectator qubits is, on average, superior to the fidelity that we obtain without using them.
Therefore, we want to satisfy:
\begin{equation}
    \langle F(\phi_s) \rangle > \langle F(\phi_n) \rangle,
    \label{condition0}
\end{equation}
which in Eq. (\ref{TaylorExpansion}) is equivalent to:
\begin{widetext}
\begin{equation}
    \langle \phi_s^2 \rangle + \frac{1}{F^{(2)}(0)} \left\langle \int_0^{\phi_s} \mathrm{d} x \; (x-\phi_s)^2 F^{(3)} (x) \right\rangle
    < \langle \phi_n^2 \rangle  + \frac{1}{F^{(2)}(0)}\left\langle \int_0^{\phi_n} \mathrm{d} x \; (x-\phi_n)^2 F^{(3)} (x) \right\rangle.
    \label{condition1}
\end{equation}
\end{widetext}
While it is possible to satisfy the necessary condition above even in situations where our spectators have not helped to decrease the effective error parameters, usually we will want to further impose that:
\begin{equation}
    \left| \phi_s \right| < \left| \phi_n \right|,
    \label{condition2}
\end{equation}
a condition that is ultimately limited by the Cram\' er--Rao bound.
However, while we try to meet this condition, we also should make sure that the errors do not increase excessively.
This can be translated into a second set of conditions that are not necessary, but are sufficient to satisfy inequality (\ref{condition1}) when we impose (\ref{condition2}).
These conditions simply say that the higher-order terms of the expansion must be negligible in comparison to the second-order terms that originate condition (\ref{condition2}):
\begin{align}
    \langle \phi_s^2 \rangle & \gg \left| \frac{1}{F^{(2)}(0)} \left\langle \int_0^{\phi_s} \mathrm{d} x \; (x-\phi_s)^2 F^{(3)} (x) \right\rangle \right| , \label{suf1} \\
    \langle \phi_n^2 \rangle & \gg  \left| \frac{1}{F^{(2)}(0)} \left\langle \int_0^{\phi_n} \mathrm{d} x \; (x-\phi_n)^2 F^{(3)} (x) \right\rangle \right| . \label{suf2}
\end{align}
Do notice that, as sufficient but not necessary conditions, we do not need to respect (\ref{suf1}) and (\ref{suf2}) in order to obtain satisfactory results with spectator qubits.
However, if the feedback loop with the spectators is not improving the fidelity of the system despite condition (\ref{condition2}) being met, then changing the scheme so that (\ref{suf1}) and (\ref{suf2}) are satisfied will suffice to make the strategy work.

It is worth remarking that this kind of analysis that requires a Taylor expansion around $\phi=0$ is not adequate for metrics that do not have a derivative at $\phi=0$, such as the diamond norm, which is defined as:
\begin{equation}
    \Diamond_{\phi} \equiv \frac{1}{2} \max _{\rho}( \|(V(\phi=0) \otimes 1)[\rho]-(V(\phi) \otimes 1)[\rho]\left\|_{1}\right).
    \label{EQ2}
\end{equation}
For this reason, we opted to use the process fidelity as defined in Eq. (\ref{fid}) in our analyses.
In any case, as there is a one-to-one correspondence between the diamond norm and the process fidelity for coherent errors, an improvement in one metric translates into an improvement in the other one as well.

Finally, let us consider a simple example of a situation where forcing conditions (\ref{suf1}) and (\ref{suf2}) to be respected also makes the complete feedback loop to function.
Suppose a data qubit and a spectator qubit simultaneously suffer the same kind of overrotation, $e^{i\phi X}$.
In this situation, the necessary condition (\ref{condition0}) takes the following exact form after $N$ measurements and overrotations:
\begin{equation}
  \cos^2 (N \phi_s) > \cos^2 (N \phi_n).
  \label{eq:example1}
\end{equation}
If a measurement is performed after each overrotation, the best average estimate we can find for $\phi$ is given by $1/(2\sqrt{N})$, according to the Cram\' er--Rao bound and Eq. (\ref{eq:Fisher}).
A simple way of satisfying (\ref{eq:example1}) is by restricting the arguments of the cosines to the interval $[-\pi/2,\pi/2]$ and then imposing condition (\ref{condition1}).
However, $\phi_s$ will only be smaller than $\phi_n$ if we perform a number of measurements that is sufficiently large to have a clear estimate of $\phi_n$ -- i.e., if $1/(2\sqrt{N}) < | \phi_n |$.
By this point, however, the increase in $N$ may have brought the angles out of the $[-\pi/2, \pi/2]$ range, in which case the necessary condition (\ref{eq:example1}) may no longer be satisfied by simply reducing the value of $\phi_s$.
In particular, this translates into a violation of the sufficient condition (\ref{suf1}).
Instead, we have (see Section I of the Supplementary Material for details of the derivation):
\begin{equation}
    \left| \frac{1}{F^{(2)}(0)} R_2(\phi_s) \right| = \left| \frac{1}{4N} + \frac{\cos (\sqrt{N})-1}{2N^2} \right|,
    \label{linearlinear}
\end{equation}
whose right-hand side approaches $\phi_s \sim 1/\sqrt{2N}$ as $N$ grows, which causes a violation of the condition.

As this is a sufficient condition, its violation at this point can be seen as merely incidental to the failure of the scheme.
Nevertheless, satisfying the sufficient conditions is all that is necessary to turn an ineffective strategy into a successful one.
In our case, if we have quantum control strategies available that are capable to slowing down the evolution of the data qubit to a fraction $\kappa < 1$ of the speed of change of the spectator, so that it now sees an effective error parameter $\kappa \phi$, we find an easier solution to be satisfied:
\begin{equation}
    \langle \phi_s^2 \rangle \gg \kappa \left\langle \left| \frac{1}{F^{(2)}(0)} \int_0^{\phi_s} \mathrm{d} x \; (x-\phi_s)^2 F^{(3)} (\kappa x) \right| \right\rangle.
\end{equation}
As long as $F^{(3)} (\phi)$ is continuous near the origin -- a feature expected for coherent errors -- the right-hand side of the inequality should go to zero as $\kappa\to 0$.
This means we can always find a sufficiently small value of $\kappa$ so that the sufficient conditions are satisfied.
Such a suppression $\kappa$ of the errors in the data qubit, which can be achieved via control techniques, is equivalent to making the spectator qubit more sensitive to errors.
This could be achieved by using different kinds of species of qubits for data and spectators.
Although this is a promising direction for future research, in this work we will focus on examples where the control techniques are responsible for the suppression.

Moreover, if the actual error parameter $\theta$ is small, we can alternatively suppress $\phi_s$ by making the data qubit perceive a quadratic error in $\theta$, while the error in the spectator remains linear.
In other words, an effective error proportional to $\sim (\theta^2 - \hat\theta^2)$ will be smaller than an effective error proportional to $\sim (\theta - \hat\theta)$.

In the simple example above we can make a feedback scheme work by increasing the relative sensitivity of the spectator to the noise. We will see in the results below that this kind of suppression is also useful in more realistic scenarios.
It is particularly convenient that the quadratic suppression of errors is common in many quantum control techniques ~\cite{KhodjastehPRA2009,SouzaPTRSA2012,KabytayevPRA2014,PokharelPRL2018}.

}

\subsection{Application to magnetic field noise}

A qubit precessing around an axis in the Bloch sphere due to some external coherent error source will behave in a manner that is analogous to a spin-1/2 subjected to an external classical magnetic field.
Calling this external classical field $\mathbf{B}$, the error will be described by the unitary $U(t) = e^{-i t \mathbf{B} \cdot \boldsymbol\sigma }$, where we are incorporating any constants into the magnitude of $\mathbf{B}$.

If we know the direction of the classical field $\mathbf{B}$, we can achieve perfect dynamical decoupling by applying $\pi$-pulses in a direction $\mathbf{\hat n}$ that is perpendicular to $\mathbf{B}$ \cite{ViolaPRL1999}.
If we do not keep track of the direction of $\mathbf{B}$, protection against first-order errors can still be obtained via repetitions of an XYXY sequence of $\pi$-pulses~\cite{GullionJMR1990,ViolaPRL1999}, also known as XY-4 \cite{PokharelPRL2018,HernandezPRB2018} or modified CPMG \cite{MaudsleyJMR1986}.
However, if we acquire information about the direction of $\mathbf{B}$ and rotate the $X$ and $Y$ pulses to a new plane $x^\prime y^\prime$ that is perpendicular to $\mathbf{B}$, this new tailored $\mathrm{X}^\prime\mathrm{Y}^\prime$-4 sequence will not only cancel perfectly the errors caused by a static $\mathbf{B}$, but will also be robust against small changes in the direction of the classical field.

\begin{figure}[htb]
\centering
\includegraphics[width=\columnwidth]{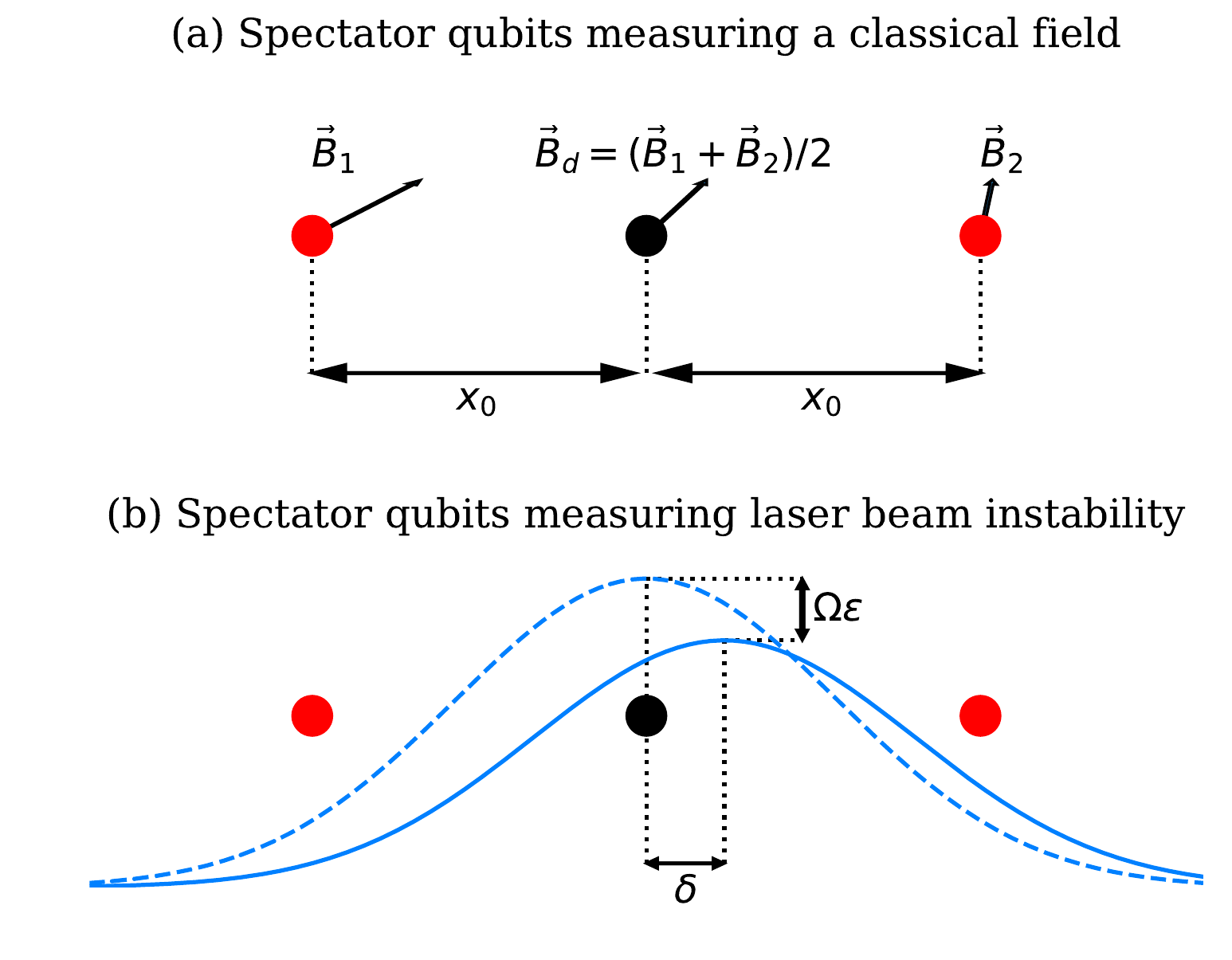}
\caption{
 \textbf{Diagrams of two spectator qubit applications.}
    In both, we assume an equal distance $x_0$ between spectators (red) and data qubit (black). In (a), a classical field is assumed to vary linearly in the position coordinate, so the field in the data ($\mathbf{B_d}$) can be estimated as the average of the field in the equidistant spectators ($\mathbf{B_1}$ and $\mathbf{B_2}$). In (b), a laser beam has its ideal Gaussian profile (dashed) changed into an actual beam (solid), which is characterized by the error parameters $\delta$ and $\varepsilon$.
}
\end{figure}

By placing spectator qubits around the data, as depicted in Fig. 2(a), we can detect drifts in the direction of a magnetic field.
We measure the components of $\mathbf{B} = B_x \mathbf{\hat x} + B_y \mathbf{\hat y} + B_z \mathbf{\hat z}$, by suppressing the undesirable parts of the qubit evolution~\cite{SekatskiNJP2016} via dynamical decoupling -- a process that can be extended to the spectroscopy of non-unitary errors as well~\cite{BylanderNP2011} and intermediate situations that involve both kinds of errors~\cite{HernandezPRB2018}.
To achieve this, we measure one component at a time, applying $\pi$-pulses in the direction that we want to measure, and preparing and measuring the spectator qubit in two distinct bases that are not eigenvectors of the pulses applied.
Meanwhile, the data qubit must undergo a dynamical decoupling that suppresses the linear terms of all the components of the magnetic field.

\begin{figure}[htb]
\centering
\includegraphics[width=\columnwidth]{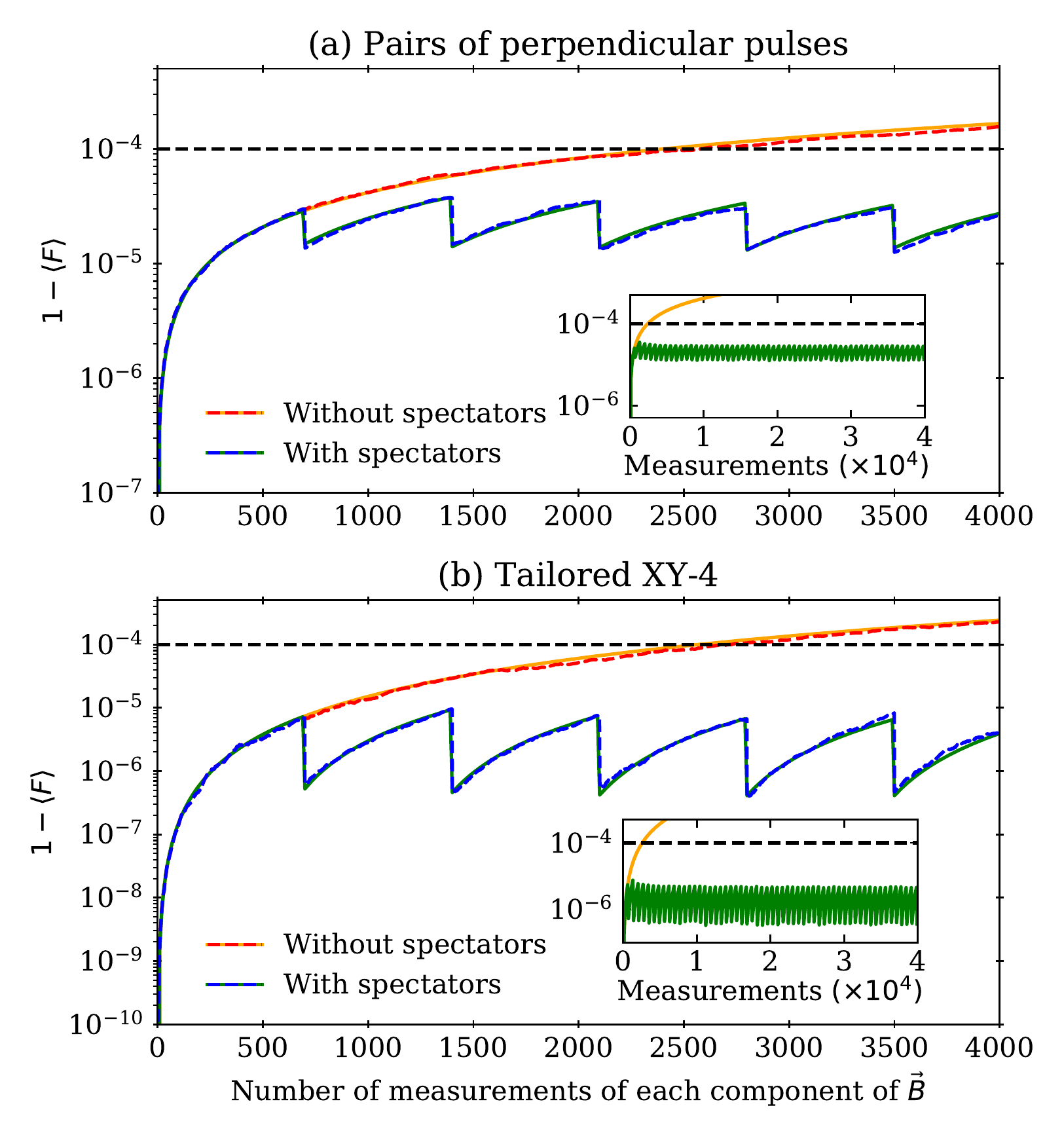}
\caption{
 \textbf{Classical field error with and without spectator qubits.}
    Here we show the average process fidelity per sequence of four $\pi$-pulses spaced by a period $\tau$, calculated numerically and averaged over 1000 runs (dashed), analytically (solid for the case without spectator qubits), and semi-analytically (solid for the case with spectator qubits) when we apply (a) just pulses perpendicular to the direction of the field; (b) a tailored XY-4 sequence where the $xy$-plane is chosen so that it is perpendicular to the magnetic field.
    Insets show the long term behavior of the fidelity, where the spectators stay indefinitely below the threshold. We assume the $\pi$-pulses to be instantaneous.
}
\end{figure}

In Fig. 3, we compare the process fidelity (\ref{fid}) for the case where we maintain the initial calibration with the case where the spectator qubits are used for recalibration. Spectator qubits are able to keep $1-\langle F \rangle$ below a $10^{-4}$ threshold after the non-recalibrated system has crossed it.
Although some codes have thresholds of the order of $1\%$~\cite{KnillNature2005,RaussendorfPRL2007,RaussendorfNJP2007,FowlerPRA2009}, a more strict threshold would allow fault-tolerance using fewer resources.
We consider the dynamical decoupling via sequences of pairs of pulses perpendicular to the direction of the magnetic field and also the tailored $\textrm{X}^\prime \textrm{Y}^\prime$-4 sequence.
In both cases, the spectator qubits stabilize at a level that remains indefinitely below the threshold.

\subsection{Application to laser beam instability}

In ion trap quantum computers, the laser beams used to drive gates, cool ions, and detect states can suffer from common calibration issues such as beam pointing instability and intensity fluctuations \cite{HaeffnerPR2008}.
Moreover, they can cause crosstalk, rotations on neighboring qubits that occur when the laser beam overlaps with more qubits than the one being addressed.
In principle, the amplitude and pointing instability can be probed by measuring the neighbors \cite{QianPRA2019}, although in practice a series of such measurements can affect other qubits in the chain, creating an additional source of errors which we discuss further in the next section.
If we assume the system allows non-disruptive measurements of single qubits, two spectators closely surrounding a data qubit become a possible way of assessing laser beam miscalibrations, as depicted in Fig. 2(b).

Variations in the amplitude of the laser beam change the Rabi frequency $\Omega$ by an amount $(1-\varepsilon)$.
Moreover, small errors in the direction of the laser beam are responsible for underrotations.
Assuming a Gaussian form for the laser beam, as illustrated in Fig. 2(b), a small pointing displacement of $\delta$ results in a quadratic change in the amplitude $\Omega$ of the laser beam affecting the data qubit:
\begin{equation}
    \Omega \left(1 - \varepsilon \right) e^{-\delta^2} \approx \Omega \left(1 - \varepsilon \right) \left( 1- \delta^2 \right).
    \label{dataGauss}
\end{equation}
At a distance $\pm x_0$ from the center of the Gaussian, the spectator qubits sense a change in amplitude that is linearly proportional to the pointing displacement $\delta$:
\begin{equation}
    \Omega \left(1 - \varepsilon \right) e^{-(\pm x_0 - \delta )^2}
    \approx \frac{\Omega}{c}  \left(1 - \varepsilon \right)  \left( 1 \pm 2 \delta \sqrt{\ln c} \right),
    \label{specGauss}
\end{equation}
where $c = e^{x_0^2}$.
This allows the spectator qubits to be sensitive enough to estimate $\delta$ before this pointing error grows too much in the data qubit.
For $\varepsilon$, the problem of having linear errors both in the data and in the spectator qubits can be overcome by applying composite pulse sequences such as SK1~\cite{MerrillACP2014}.
This kind of sequence reduces the effect of the error in the data qubit to a higher order, while preserving the linear effect on the spectator qubits.

\begin{figure}[htb]
\centering
\includegraphics[width=\columnwidth]{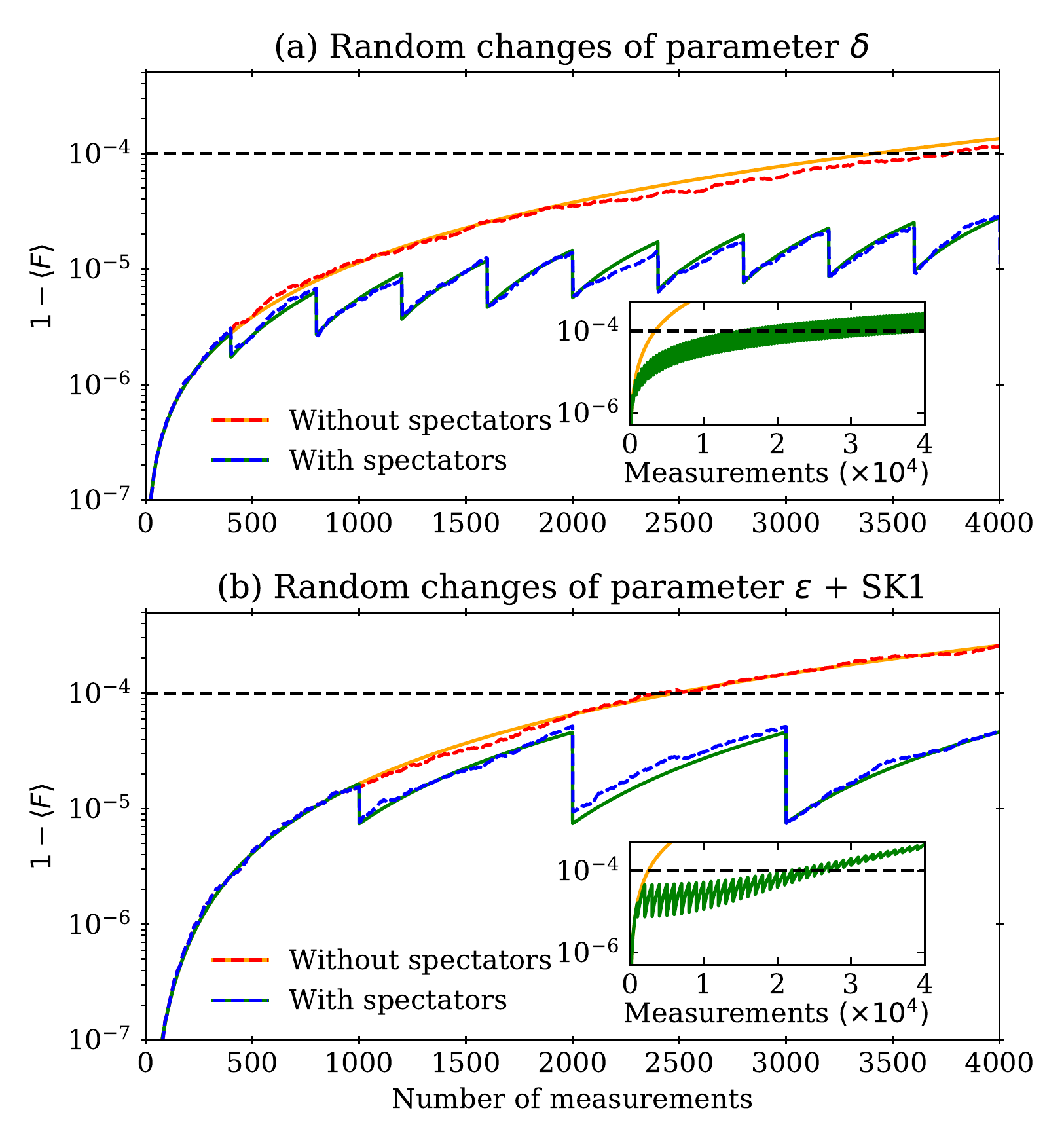}
\caption{
\textbf{Beam instability errors, with and without spectator qubits.}
    This is the evolution in time of the average process fidelity per gate for fluctuating parameters (a) $\delta$; and (b) $\varepsilon$.
    Discontinuities in the average solution over 1000 numerical runs (dashed) and the analytical approximate solutions (solid) represent points where there is a recalibration.
    Insets show analytical solutions for the longer time scales, showing the point where the recalibrated systems cross the threshold.
}
\end{figure}

The process fidelity for the laser instability with and without recalibration with spectator qubits is shown in Fig. 4.
The improvement in fidelity means that we are acquiring information fast enough to be able to recalibrate the system before the errors become too large.
If spectator qubits and data both were subjected to an error linearly proportional to $\varepsilon$, the recalibration would not be able to keep the errors under the same threshold for the same values of the parameters, as can be seen in Fig. 5.
It is therefore crucial to choose a measurement strategy that balances the rate of acquisition of information and the rate with which the errors increase.  We define $\tau_\text{nospec}$ to be the time when $\langle F_\text{nospec} \rangle$ crosses this threshold and $\tau_\text{spec}$ when  $\langle F_\text{spec} \rangle$ crosses the threshold.
In Fig. 6, we show which combinations of random walk parameters and measurements per spectator cycle $M$ are still capable of providing an effective recalibration mechanism.
{Fig. 6 also functions as a control landscape and a map that shows the frequency of updates that is adequate for a given order of magnitude of the error.
 Using our estimate of the order of magnitude of the error obtained from the initial calibration, we can find the adequate frequency of update of the control method by looking for the blue regions below the solid black line.
}

\begin{figure}[htb]
\centering
\includegraphics[width=\columnwidth]{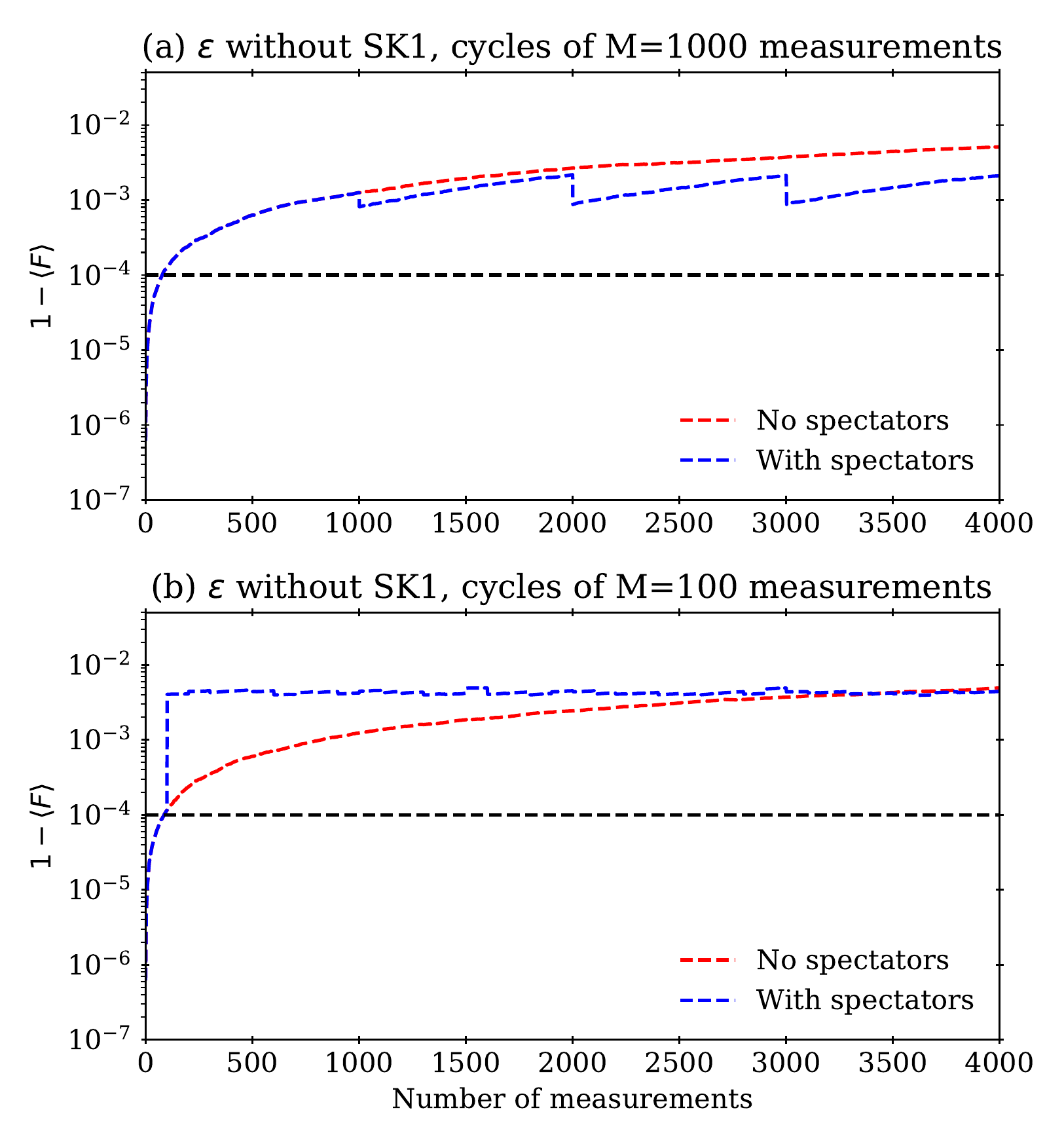}
\caption{
\textbf{Example of a frustrated spectator qubit scheme.}
        This numerical simulation is averaged over 1000 runs for the same situation as Fig. 4 (b), but with the SK1 turned off, so that the effect of the error parameter $\varepsilon$ is linear in both spectators and data qubits.
        Under these circumstances, it is never possible to keep the average error under the $10^{-4}$ threshold: either (a) we take too long to use the information from the spectator qubits and the correlation is already lost, or (b) we update before sufficient Fisher information is available, causing further miscalibration of the system.
}
\end{figure}

\begin{figure*}
\centering
\includegraphics[width=\textwidth]{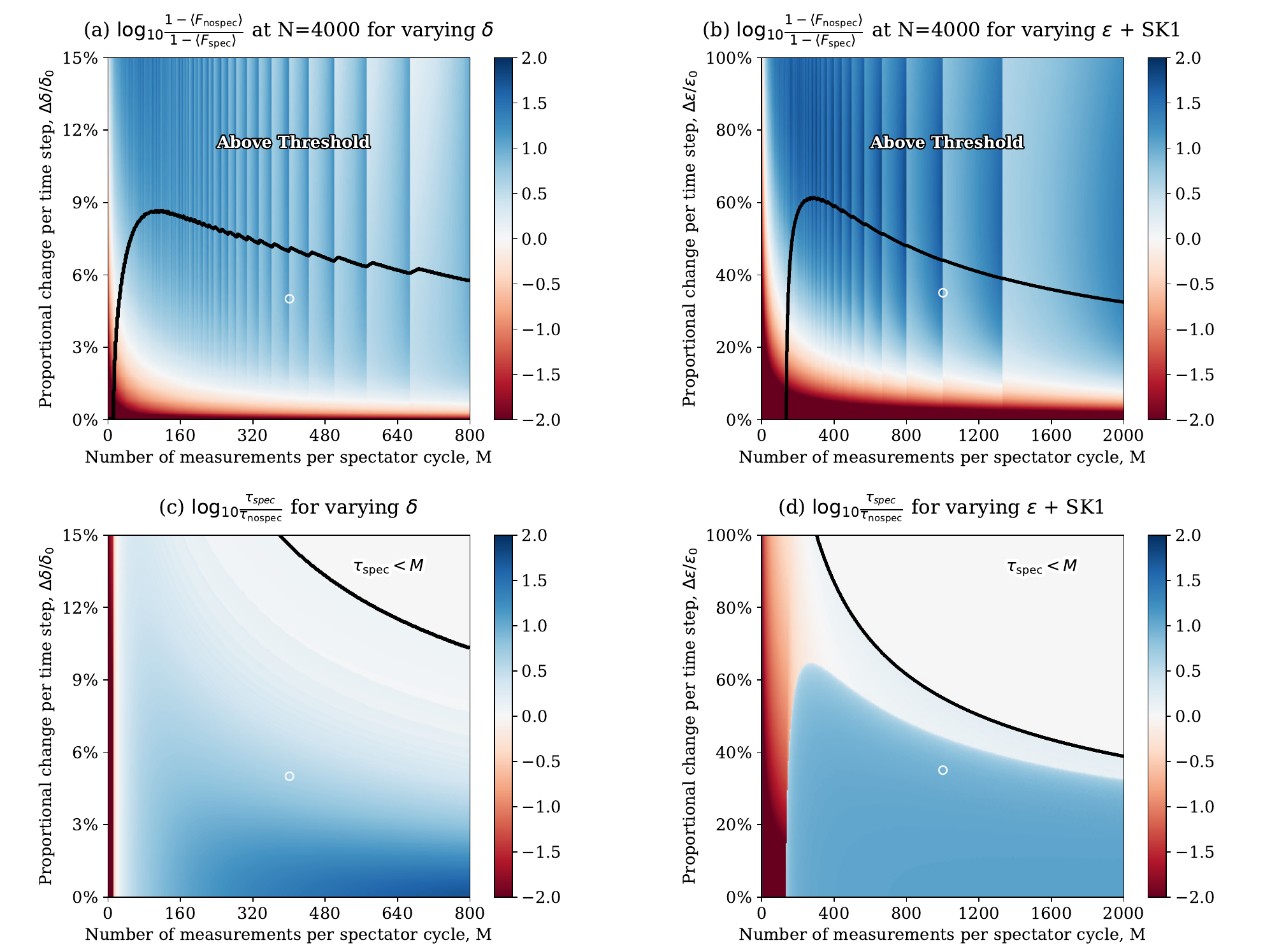}
\caption{
\textbf{Control landscapes for beam instability errors.}
    These are heat maps in base--10 logarithm of the: (a,b) ratio of $1-\langle F \rangle$ with and without spectator qubits at fixed time $N=4000$; and (c,d) ratio between the points where the $10^{-4}$ threshold is crossed, with and without spectator qubits.
    These control landscapes are heat maps showing the effect of spectator qubits for different values of measurements per cycle ($M$) and step size.
    Initial values of the error parameters and ion distance are the same as in Fig. 4, and the values of $\Delta \varepsilon$, $\Delta \delta$, and $M$ that correspond to those from Fig. 4 are marked by the white circle.
    Blue regions represent settings where the spectators improve either (a,b) the fidelity; or (c,d) time it stays below the $10^{-4}$ threshold.
    Regions above the black curve represent situations where: (a,b) the system with spectators cross the $10^{-4}$ threshold for the process fidelity; or (c,d) where the $10^{-4}$ threshold is crossed before there has been time to complete the first spectator cycle.
    Spectator qubits perform worse (red areas) when very few measurements are performed before updating (left extremity of the graphs), or when the rate of change is so small that not recalibrating is a better strategy, as in the bottom of (a) and (b).
    Discontinuities along the $x$-axis in (a) and (b) correspond to situations where the end of a spectator cycle occurs at the point $N=4000$, and are analogous to the discontinuities seen in Fig. 4.
}
\end{figure*}

\section{Discussion}

We have shown that spectator qubits are capable of recalibrating an error reduction scheme for coherent errors with a precision that is only limited by the Fisher information available and by our capacity of slowing down the rate with which the error in the data qubit changes with time. {   In this work, both spectator and data qubits were equally sensitive to noise, and we had to use quantum control methods to slow down the error accumulation on data qubit. One benefit of such a strategy is that we can dynamically allocate spectator qubits \cite{GuptaARXIV2019}, and use all physical qubits in our system as spectator or data qubits. But we can sacrifice this advantage to construct (or choose) spectator qubits that are more sensitive to a particular type of noise than our data qubits. Such a strategy eliminates the need for quantum control methods that were used to slow down errors on data qubit. One example is to use multi-species ions in ion-trap architecture \cite{BrownPRA2019}. We can use Zeeman qubits (first-order magnetic field sensitive) as our spectator qubits and hyperfine qubits (first-order magnetic field insensitive) as our data qubits for dealing with magnetic field noise.             }

In the case of the spectator qubits used to calibrate dynamical decoupling, in Fig. 3, we have seen the fidelity remain under the threshold for an indefinite period.
We believe this is possible because the error in this setting depends mainly on the angle between the classical field and the pulses, a parameter whose value is not allowed to grow indefinitely.

For laser beam instability, the insets in Fig. 4 show that even when we are using spectator qubits, the average gate fidelity crosses over the threshold at a later time. We believe this is because the error parameters become very large as the random walk is unbounded. This contrasts with the magnetic field parameters, whose random walk was bounded. When the error parameters become very large ($\varepsilon$, $\delta$ $>$ 1), the data qubit (error is quadratic) becomes more sensitive to the error than the spectator qubits (error is linear). One possible way to fix this is to include an external classical controller that restricts the maximum variance of the fluctuating error parameters and prevents the crossing of the threshold.

It is worth noting for the laser beam instability case that
although we have simulated the fidelity of a single gate ($\sigma_{x}$) due to miscalibration, it is straightforward to extend our approach to an arbitrary computation. We can do this by interleaving cycles composed of gates that we want to calibrate on data qubits and spectator qubit measurements between gates of the algorithm.

{

 When attempting to put the beam instability feedback loop into practice, it may become important to take into account the additional errors that can be caused by measuring neighboring qubits.
 However, in this setting where we attempt to improve the process fidelity per gate, the penalty for the measurement errors does not accumulate, rather impacting the performance of the gates individually.
 If a measurement error is modeled by an incoherent process that occurs with probability $p$, this will count as an incoherent channel that occurs after the gate and reduces the probability of no error occuring by a fraction $1-p$.
 This new background error will reduce overall gate fidelity and in order to maintain gates below the threshold error, it is critical to maintain an even better calibration. In effect this will lower the threshold that we have to reach by a fixed amount, but the overall analysis of the problem remains the same.
 When the measurement error occurs during an SK1 sequence, the interaction between the series of gates and the incoherent channel will be more complicated, but the overall effect will still be that an incoherent error spoils the state of the system with probability $p$, having the overall effect of lowering the threshold.
}

The possibilities of applications of spectator qubits are not limited to the two coherent errors that we simulated above.
Protection against magnetic fields, for example, besides being relevant to ion traps and nuclear spin qubits, could be extended to detection and dynamical decoupling of a classical external electric field $\mathbf{E}$ for qubits that are instead sensitive to electric fields, such as antimony nuclei~\cite{AsaadARXIV2019}.

In future full-fledged quantum computing systems, spectator qubits will enable error rates below the threshold for fault-tolerance for longer times than systems without spectator recalibration. This will allow for longer quantum computations.
For near and medium-term applications, however, enhancements would be required in order to reduce the prohibitive number of measurements necessary to obtain a reliable estimate of the change in the calibration.
It would be particularly desirable to implement small corrections in the calibration after fewer measurements, possibly assuming some prior knowledge of how the calibration changes, or a specific biased drift of the error parameters.
These could be combined with other venues for improvement, such as using Bayesian learning protocols~\cite{GuptaPRApplied2018,GuptaARXIV2019} to make more accurate previsions of future evolution of error parameters or to implement adaptive measurements~\cite{GrenadeNJP2012,DemkowiczPRX2017},  and using entangled states~\cite{EldredgePRA2018}, many-body Hamiltonians~\cite{DemkowiczPRX2017}, quantum codes~\cite{ZhouNature2018}, or optimal control~\cite{LiuPRA2017} to maximize the information available.

\section{Methods}

\subsection{Error model}

In our numerical, analytical and semi-analytical simulations for Figs. 3, 4, 5, and 6,
the error parameters $\theta$ are assumed to start at a fixed value $\theta_0$ and fluctuate in time according to a random walk with unbiased Gaussian steps of average size $\Delta \theta$, so that the probability of it having a value $\theta_N$ after $N$ steps will be:
\begin{equation}
    p (\Theta = \theta|\Theta_0 = \theta_0 ) =
    \mathcal{N} (\theta_0, N(\Delta \theta)^2; \theta),
    \label{prob_random_walk}
\end{equation}
where the random variable $\Theta_n$ gives the value of the error parameter after $n$ steps, and $\mathcal{N} (\mu, \sigma^2; x)$ is the normal distribution with mean $\mu$ and variance $\sigma^2$:
\begin{equation}
    \mathcal{N} (\mu, \sigma^2; x) =
    \frac{1}{\sqrt{2\pi \sigma^2}}
    \exp \left\{
        - \frac{(x-\mu)^2}{2 \sigma^2}
    \right\}.
    \label{normal_dist}
\end{equation}
Suppose that, given an actual value of the error parameter $\theta$ and an estimate $\vartheta$, we know the expression of the process fidelity per gate, $F(\theta,\vartheta)$.
Then, if the parameter drifts in time but our estimate is not updated, we can use the probability distribution of the random walk to find the average fidelity per gate after $N$ steps when spectators are not recalibrating the system, $F_\text{nospec}$:
\begin{equation}
    \langle F_\text{nospec} \rangle = \int \mathrm{d} \theta\; p (\Theta_N = \theta|\Theta_0 = \theta_0) F(\theta,\vartheta).
    \label{no_spectator}
\end{equation}
This expression can be analytically calculated for all the application above (see Supplementary Section III).

If spectator qubits are present, the estimate $\vartheta$ is updated after every cycle of $M$ measurements.
After the $k$th cycle of measurements of the spectators, the next estimate is obtained via an estimator $\theta^\text{(est)}_{kM}$ that consists of the average of the error parameter sampled at the previous $M$ steps of the random walk:
\begin{equation}
    \Theta^\text{(est)}_{kM} = \frac{\Theta_{(k-1)M+1}+ \Theta_{(k-1)M+2} + \ldots + \Theta_{kM}}{M},
    \label{estimator}
\end{equation}
where we assume that the parameters change sufficiently slowly so that $\theta$ has a precisely defined value  during each measurement.
For this reason, the variance of the Gaussian in Eq. (\ref{prob_random_walk}) can always be rescaled so that the number of steps of the random walk matches the number of measurements.

The probability distribution of the estimator will be a Gaussian, as this is a random variable consisting of the average of the Gaussian random variables $\Theta_{kM+n}$:
\begin{equation}
    p(\Theta^\text{(est)}_{kM} = \theta^\text{(est)}_{kM} | \Theta_{kM} = \theta_{kM} )
    = \mathcal{N} (\mu_k, \sigma^2_k; \theta^\text{(est)}_{kM}).
\end{equation}
Therefore, the probability distribution will be entirely characterized by the two cumulants that can be calculated from Eq. (\ref{estimator}), which are (see Supplementary Section II) the mean $\mu_k$:
\begin{equation}
    \mu_k = \theta_{kM} + \frac{M-1}{kM} \frac{\theta_0-\theta_{kM}}{2},
    \label{mean}
\end{equation}
and the variance $\sigma_k^2 $:
\begin{equation}
    \sigma^2_k =
    \frac{M-1}{3} \frac{4kM - 3M - 3k + 2}{4kM} (\Delta \theta)^2.
    \label{variance}
\end{equation}

Finally, the measured value $\bar\theta$ may differ from the estimator, according to the lower limit of the Cram\' er--Rao bound (\ref{CramerRao}), by an amount that corresponds to the inverse of the Fisher information $f_\theta$ times the number of measurements:
\begin{equation}
    p(\bar \Theta = \bar\theta | \Theta^\text{(est)}_{kM} = \theta^\text{(est)}_{kM} )
     = \mathcal{N} (\theta^\text{(est)}_{kM}, (M f_\theta)^{-1}; \bar\theta).
\end{equation}

Given these probability distributions for $\theta_{kM}$, $\theta_{kM}^\text{(est)}$, and $\bar\theta$, the average fidelity $N$ steps after the $k$th spectator cycle, which we call $ F_\text{spec}$,  can be calculated from the average fidelity for a fixed calibration given in Eq. (\ref{no_spectator}):
\begin{multline}
    \langle F_\text{spec} \rangle =
    \left. \int \mathrm{d} \theta_{kM} \; \right\{ p (\Theta_{kM} = \theta_{kM} |\Theta_0 = \theta_0 ) \\
    \times \int \mathrm{d} \theta^\text{(est)}_{kM} \; \left[
    p(\Theta^\text{(est)}_{kM} = \theta^\text{(est)}_{kM} | \Theta_{kM} = \theta_{kM} ) \right. \\
    \left. \left.
    \times \int \mathrm{d} \bar\theta \;
     p(\bar \Theta = \bar\theta | \Theta^\text{(est)}_{kM} = \theta^\text{(est)}_{kM} )
    \langle F_N(\theta_0, \bar\theta) \rangle_0 \right] \right\}.
    \label{complete}
\end{multline}
Using the assumption that the error parameters are small, we solved the triple integrals analytically for $\varepsilon$ and $\delta$ (see Supplementary Section IV) and numerically for the magnetic field case.

\subsection{Magnetic field noise}

In the simulation of the dynamical decoupling of a magnetic field, we assumed the field gradient to be linear, so that measurements in two spectators are sufficient to determine the field in the data qubit.
We choose the $\mathbf{\hat z}$ axis to coincide with the initial direction of the magnetic field.
Using $\tau$ to denote the time spacing between instantaneous $\pi$-pulses, we choose the initial value $\mathbf{B}_1$ of the magnetic field in one of the spectators to satisfy $\tau \mathbf{B}_1 = 2 \cdot 10^{-3}\; \mathbf{\hat z}$ when the data qubit undergoes a 2-pulse sequence, and $\tau \mathbf{B}_1 = 3.8 \cdot 10^{-2}\; \mathbf{\hat z}$ when it undergoes the tailored XY-4 sequence.
The second spectator is assumed to experience initially half of the value of this magnetic field ($\mathbf{B}_2 = \mathbf{B}_1/2$).

Each component of the magnetic field was assumed to perform an independent random walk, with steps of different size.
We choose standard deviations $\Delta B_x/B_{1,x}= 3\%$, $\Delta B_y/B_{1,y}= 2\%$, $\Delta B_z/B_{1,z}= 1\%$ for each random walk.
These components are then assessed separately and sequentially in the spectator qubits, which is done by preparing and measuring the spectator in eigenbases of two distinct Pauli matrices that are perpendicular to the component of $\mathbf{B}$ that we want to measure. { There is no need for spectator qubit re-initialization after each measurement as measurement in a particular Pauli basis prepares the spectator qubit in an eigenbases of that Pauli operator.   }
The other components of $\mathbf{B}$ are decoupled by applying a sequence of $\pi$-pulses to the spectators between each measurement, so that we can approximate our estimates of $B_x$, $B_y$, and $B_z$ by:
\begin{align}
    \tilde B_x = &
        - \arcsin \left( \frac{ \left\langle 0 \right| U^\dagger (n\tau) \sigma_y U(n\tau) \left| 0 \right\rangle }{4n\tau} \right), \\
    \tilde B_y = &
        \arcsin \left( \frac{ \left\langle 0 \right| U^\dagger(n\tau) \sigma_x U(n\tau) \left| 0 \right\rangle }{4n\tau} \right),  \\
    \tilde B_z = &
        \arcsin \left( \frac{ \left\langle + \right| U^\dagger(n\tau) \sigma_y U(n\tau) \left| + \right\rangle }{4n\tau} \right),
\end{align}
where $\left\langle \psi \right| U^\dagger(n\tau) \sigma_i U(n\tau) \left| \psi \right\rangle$ represents the averages of a measurement of $\sigma_i$ in a system prepared at a state $\left| \psi \right\rangle$ and left to evolve for a time $n \tau$.
The number $n$ of $\pi$-pulses before each measurement was chosen as $20$ for spectators aiding the perpendicular 2-pulse sequence, and $4$ for spectators whose information was used to tailor a XY-4 sequence.
After $M=700$ measurement cycles, we use the new estimate of the direction of $\mathbf{B}$ to update { our dynamical decoupling control parameters.  For the pairs of perpendicular $\pi$-pulses, we make the pulse direction perpendicular to $\mathbf{B}$. We estimate the plane that is normal to $\mathbf{B}$, find two perpendicular pulses in that plane, and use them as our tailored XY-4 sequence. We repeat this process throughout the length of our computation.  }

\subsection{Laser beam instability}

For spectator qubits used to reduce the underrotation caused by pointing instability of the laser beam, we simulate a series of $\sigma_x$ gates applied to the data qubit and assume the presence of either a fluctuating parameter $\delta$ or $\varepsilon$.
The $\delta$ parameter is assumed to start the random walk at $\delta_0 = 0.02$ and proceed with Gaussian steps of standard deviation $\Delta \delta/\delta_0 = 5\%$, while the $\varepsilon$ starts at $\varepsilon_0 = 0.002$ and proceeds with steps of standard deviation $\Delta \varepsilon/\varepsilon_0 = 35\%$.
We assume an initial calibration that allows us to estimate $\delta$ to a precision $\bar \delta/\delta_0 = 99 \%$, and $\varepsilon$ to a precision $\bar\varepsilon/\varepsilon_0 = 75 \%$.

The $\delta$ errors naturally cause a greater effect in the spectators than in the data, as can be seen from Eqs. (\ref{dataGauss}) and (\ref{specGauss}), where the difference is between a linear and a quadratic dependence.
After four $\sigma_x$ gates, we measure the spectator qubits, which we assume to be at a distance  $x_0=(\ln 12)^{1/2}$ from the data ($x_0^2$ is measured in units of twice the variance).
To maximize the Fisher information for uncorrelated probes, the two spectator qubits are prepared and measured in an eigenstate of $\sigma_z$.
The averages $\langle \sigma_z \rangle_1$, $\langle \sigma_z \rangle_2$ of the $M$ measurement results in spectator qubits 1 and 2 are then used to estimate $\delta$ according to:
\begin{equation}
    \bar\delta = \frac{1}{2x_0} \frac{ \arccos(\left\langle \sigma_z \right\rangle_1) - \arccos(\left\langle \sigma_z \right\rangle_2) }{\arccos(\left\langle \sigma_z \right\rangle_1) + \arccos(\left\langle \sigma_z \right\rangle_2)},
\end{equation}
which follows from Eq. (\ref{specGauss}).
As only $\delta^2$ affects the data, the sign of our estimate of $\delta$ is irrelevant.
After $M=400$ repeated measurements, we build sufficient confidence in our estimate $\bar\delta$ so that, for future gates, we adjust the Rabi frequency to $\Omega/(1-\bar\delta^2)$ { using our classical control setup }  to compensate for the pointing instability.

As the parameter $\varepsilon$ is linear in all qubits, we apply an SK1 composite pulse sequence~\cite{BrownPRA2004} to slow down the error accumulation in the data qubit.
Measurements on the spectator qubits -- assumed to be at a distance $x_0= (\ln 1.8)^{1/2}$ from the data (where $x_0^2$ is measured in units of twice the variance) -- are performed after each regular $\sigma_x$ gate is applied, but before the application of the second and third pulses of the SK1 sequence.
The value of $\varepsilon$ is then estimated from the measurement results of $\langle \sigma_z \rangle_1$, $\langle \sigma_z \rangle_2$:
\begin{equation}
    \bar\varepsilon = 1- e^{x_0^2} \frac{ \arccos(\left\langle \sigma_z \right\rangle_1) + \arccos(\left\langle \sigma_z \right\rangle_2) }{2\pi/(1-\tilde\varepsilon)},
\end{equation}
where $\tilde\varepsilon$ is the previous estimate of $\varepsilon$.
After $M=1000$ measurements, we update the Rabi frequency to $\Omega/(1-\bar\varepsilon)$ to compensate for the errors.

\section*{Data and code availability}
\noindent
The data and code that support the findings of this study are available from the corresponding authors upon reasonable request.

\section*{Acknowledgments}
\noindent The authors would like to thank Natalie Brown, Dripto Debroy, Shilin Huang, Pak Hong Leung, Iman Marvian, Michael Newman, Gregory Quiroz, and Lorenza Viola for helpful discussions.
Numerical simulations were performed on the Duke Compute Cluster (DCC).
This work was supported by the ARO MURI grant W911NF-18-1-0218.

\vspace{1em}

\noindent This is a pre-print of an article published in npj Quantum Information. The final authenticated version is available online at: https://doi.org/10.1038/s41534-020-0251-y.

\section*{Competing interests}
\noindent
The authors declare that there are no competing interests.

\section*{Author contributions}
\noindent
K.R.B. conceived the idea of this paper and directed the project.
S.M. and L.A.C. modelled the problem and obtained the simulated results. S.M., L.A.C., and K.R.B. analyzed the data and prepared the manuscript.

\end{document}


\title{Supplementary information for: \\ Real-time calibration with spectator qubits}

\author{Swarnadeep Majumder}
\affiliation{Departments of Electrical and Computer Engineering, Chemistry, and Physics, Duke University, Durham, NC 27708 USA}%
\author{Leonardo Andreta de Castro}\email{leoadec@leoad.ec}
\affiliation{Departments of Electrical and Computer Engineering, Chemistry, and Physics, Duke University, Durham, NC 27708 USA}%
\affiliation{Present address: Q-CTRL Pty Ltd, Sydney, NSW Australia}
\author{Kenneth R. Brown}
\affiliation{Departments of Electrical and Computer Engineering, Chemistry, and Physics, Duke University, Durham, NC 27708 USA}%

\date{9 February 2020}

\maketitle

In this document, we present additional details of the calculations used in the article.
In Supplementary Section I, we explain in more detail the equations for the case where both data qubit and spectator qubit are subject to linear errors.
Supplementary Section II has the derivation of the probability distribution for the estimator of an error parameter undergoing a random walk.
Both Supplementary Sections III and IV contain the analytical solutions for the fidelities in the applications for the cases where a spectator qubit is not present, and when it is, respectively.
Finally, Supplementary Section V describes the difficulties of applying the Robust Phase Estimation (RPE) for a parameter that is changing in time.

\section{Details of the linear vs. linear case}

To derive Eq. (16), we take into account that the gate fidelity $F(\phi)=\cos^2(N\phi)$ for $N$ overrotations of an amount $\phi$ has the following second and third derivatives:
\begin{align*}
    F^{(2)} (\phi)  = & - 2 N^2 \cos(2N\phi), \\
    F^{(3)} (\phi)  = & 4 N^3 \sin(2N\phi).
\end{align*}

\begin{figure}[htb]
    \centering
    \includegraphics[width=\columnwidth]{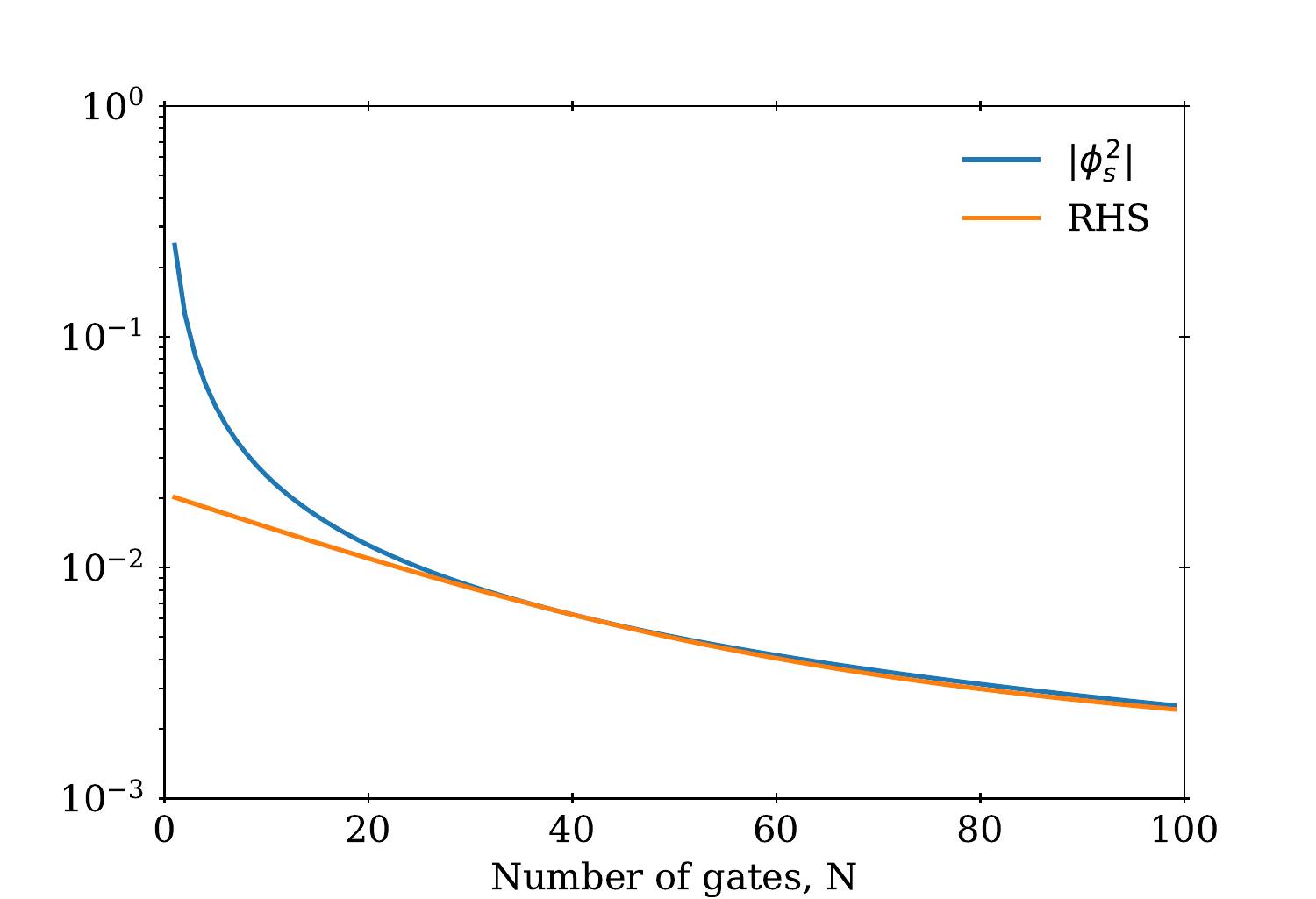}
    \caption{$|\phi_s^2|$ and right-hand side (RHS) of condition (12) for $\phi_s=(4N)^{-1}$, as in Eq. (16). After $\sim 10$ gates, the two sides are already of the same order of magnitude, and condition (12) is no longer satisfied.}
    \label{fig:appendix}
\end{figure}

Replacing these expressions on the right-hand side of condition (12), we find:
\begin{equation*}
    \left| \frac{1}{F^{(2)}(0)} R_2(\phi_s) \right| =
    \left|
        2N \int_0^{\phi_s} \mathrm{d} x\;
        (x-\phi_s)^2 \sin(2Nx)
    \right|,
\end{equation*}
where $R_2(\phi)$ is two times the Lagrange remainder:
\begin{equation}
    R_2(\phi) =
    \int_0^\phi \mathrm{d} x \; (x-\phi)^2 F^{(3)} (x).
\end{equation}
After integration, this becomes:
\begin{equation*}
    \left| \frac{1}{F^{(2)}(0)} R_2(\phi_s) \right| =
    \left|
        \phi_s^2
        + \frac{\cos(2N\phi_s)-1}{2N^2}
    \right|.
\end{equation*}
For large $N$, the term of the order of $\phi_s^2$ predominates.
If $\phi_s^2=(4N)^{-1}$, this expression becomes the same as Eq. (16).
How these terms become similar as $N$ grows can be seen in Fig. \ref{fig:appendix}.

\section{Probability distribution of the estimator}
\label{app:estimator}

In Eq. (23), we defined a random variable $\Theta^\text{(est)}_{kM}$ that was the average of the Gaussian random variables $\Theta_{kM}$, $\Theta_{kM-1}$, $\ldots$, $\Theta_{kM-(M-1)}$.
The probability distribution of this random variable will therefore also be a Gaussian, and can be fully characterized by its first two cumulants, $\mu_k$ and $\sigma^2_k$.
To find these two, we just need the means, variances and covariances of the individual Gaussian random variables $\Theta_k$, as can be seen from the following:
\begin{align*}
    \mu_k = & \frac{1}{M} \sum_{n=0}^{M-1} \langle \Theta_{kM-n} \rangle , \\
    \sigma_k^2 = & \frac{1}{M^2} \sum_{n, q=0}^{M-1} \left[ \langle \Theta_{kM-n} \Theta_{kM-q} \rangle - \langle \Theta_{kM-n} \rangle \langle \Theta_{kM-q} \rangle \right].
\end{align*}
Therefore, we just need the probability distribution for the individual positions in the random walk.
As long as space is isotropic, the conditional probability distribution for $\Theta_{kM-n}$ will be proportional to the product of two random walks starting at $\Theta_0 = \theta_0$ and $\Theta$ the extremities and ending at $\theta_{kM-n}$:
\begin{multline*}
    p(\Theta_{kM-n} = \theta_{kM-n} | \Theta_0 = \theta_0, \Theta_{kM} = \theta_{kM} ) \\
    \propto
    p(\Theta_{kM-n} = \theta_{kM-n} | \Theta_0 = \theta_0 )
    p(\Theta_{n} = \theta_{n} | \Theta_0 = \theta_{kM} ).
\end{multline*}
After normalization, we find a Gaussian distribution of the form given by Eq. (21):
\begin{multline*}
    p(\Theta_{kM-n} = \theta_{kM-n} | \Theta_0 = \theta_0, \Theta_{kM} = \theta_{kM} ) \\
    \;\;\; = \mathcal{N} \left( \theta_{kM-n};
    \theta_{kM} + n \frac{\theta_0-\theta_{kM}}{kM},
    \frac{n(kM-n)}{kM} (\Delta \theta)^2 \right).
\end{multline*}
The mean of the estimator $\Theta_{kM}^\text{(est)}$ can then be found from the average of the random variables $\Theta_{kM-n}$:
\[
    \mu_k = \frac{1}{M} \sum_{n=0}^{M-1} \left( \theta_{kM} +  n \frac{\theta_0-\theta_{kM}}{kM} \right),
\]
which trivially yields Eq. (25).

To find $\sigma_k^2$, we will also need the covariances between any two random variables $\Theta_{kM-n}$ and $\Theta_{kM-q}$, which can be found from the correlation function $\rho$:
\begin{multline*}
    \langle \Theta_{kM-n} \Theta_{kM-q} \rangle - \langle \Theta_{kM-n} \rangle \langle \Theta_{kM-q} \rangle \\
    = \sqrt{\frac{n(kM-n)}{kM}} \sqrt{\frac{q(kM-q)}{kM}} (\Delta \theta)^2 \rho.
\end{multline*}
The correlation function can be found from the coefficients that multiply the quadratic terms in the exponents of the joint Gaussian distribution:
\begin{multline*}
     p(\Theta_{kM-n} = \theta_{kM-n} | \Theta_0 = \theta_0 ) p(\Theta_{q} = \theta_{kM} | \Theta_0 = \theta_{kM-q} )\\
     \times
     p(\Theta_{n-q} = \theta_{kM-q} | \Theta_0 = \theta_{kM-n} )
      \\
     \propto \exp \left\{
        - \frac{kM}{(1-\rho^2)n(kM-n)} \frac{ \theta_{kM-n}^2 }{ 2 (\Delta \theta)^2 }
    \right\} \\
    \times \exp \left\{
        - \frac{kM}{(1-\rho^2)q(kM-q)} \frac{ \theta_{kM-q}^2 }{ 2 (\Delta \theta)^2 }
     \right\},
\end{multline*}
where we assumed $n\ge q$, without loss of generality.

Comparing with the known probability distributions for the random walk, we find:
\[
    \rho^2 = 1 - \frac{kM(n-q)}{n(kM-q)} = \frac{q(kM-n)}{n(kM-q)}.
\]
Then, we find a simple expression for the covariance of the individual random variables, which can be translated into the variance of the estimator:
\[
    \sigma_k^2 = \frac{1}{M^2} \sum_{n=0}^{M-1} \left[ \frac{n(kM-n)}{kM} + 2  \sum_{q=0}^{n-1} \frac{q(kM-n)}{kM}
    \right] (\Delta \theta)^2.
\]
Using Faulhaber's formulas to calculate $\sum n^2$ and $\sum n^3$, it is straightforward to recover the result from Eq. (26).

\section{Analytical expressions for the fidelity without spectator qubits}
\label{app:nospec}

\subsection{Pairs of perpendicular pulses}
A sequence of four pulses in the direction $\mathbf{\hat e_x}$ spaced by time intervals $\tau$, while a system undergoes a rotation by a classical magnetic field $\mathbf{B}$, will result in the following process fidelity:
\[
  F = \frac{1}{4} \left| \mathrm{Tr} \left\{
    \left[
        e^{- i \mathbf{B} \cdot \boldsymbol\sigma \tau}
        (\mathbf{\hat e_x} \cdot \boldsymbol\sigma)
    \right]^4
  \right\} \right|^2.
\]
The average of this fidelity can then be written in terms of components that are parallel ($B_\parallel$), and perpendicular ($B_\perp$) to $\mathbf{\hat e_x}$:
\begin{multline*}
    \langle F_\text{nospec} \rangle =
    1 - 16 \langle B_\parallel^2 \rangle \tau^2
    + \frac{16}{3} \langle B_\parallel^2 B^2 \rangle \tau^4
    + 16 \langle B_\parallel^4 \rangle \tau^4 \\
    + O(B^6\tau^6).
\end{multline*}
The extant sum of averages of components of $\mathbf{B}$ are replaced by the appropriate moments of the random walks of $B_x$, $B_y$, and $B_y$, which correspond to Gaussians centered in the initial value of $\mathbf{B}$ and with standard deviations that grow with $\sqrt{N}$.
$\mathbf{\hat e_x}$ is chosen to be proportional to the cross product of the current estimate of the magnetic field, $\mathbf{\tilde B}$ and some random vector, being therefore:
\begin{multline*}
  \mathbf{\hat e_x} =
  \frac{3}{2\tilde B} \left(
    \sqrt{1-u^2} \sin \phi \tilde B_z - u \tilde B_y
  \right) \mathbf{\hat x} \\
  + \frac{3}{2\tilde B} \left(
    u \tilde B_x - \sqrt{1-u^2} \cos \phi \tilde B_z
  \right) \mathbf{\hat y} \\
  + \frac{3}{2\tilde B} \left(
    \sqrt{1-u^2} \cos \phi \tilde B_y - \sqrt{1-u^2} \sin \phi \tilde B_x
  \right) \mathbf{\hat z},
\end{multline*}
where $u$ and $\phi$ are sampled uniformly from intervals $[-1,1]$ and $[0,2\pi]$, respectively.

\subsection{Tailored XY-4 dynamical decoupling}
Given alternated pulses in the $\mathbf{\hat e_x}$ and $\mathbf{\hat e_y}$ directions, the average process fidelity of a qubit under a classical field $\mathbf{\hat B}$ will be, after a sequence of four pulses spaced by a time $\tau$;
\[
  F  = \left|
    1 -
    8 (\mathbf{\hat e_x} \cdot \mathbf{B})^2 (\mathbf{\hat e_y} \cdot \mathbf{B})^2
    \frac{\sin^4 (B\tau)}{B^4}
  \right|^2.
\]
The unit vectors $\mathbf{\hat e_x}$ and $\mathbf{\hat e_y}$ are chosen to be perpendicular to each other and the current estimate of the magnetic field, $\mathbf{\tilde B}$.
The vector $\mathbf{\hat e_x}$ is proportional to the cross product between $\mathbf{\tilde B}$ and some random unit vector $\mathbf{\hat n}$, while $\mathbf{\hat y}$ is proportional to the cross product between $\mathbf{\tilde B}$ and $\mathbf{\hat e_x}$, so that average the process fidelity will be:
\[
  \langle F_\text{nospec} \rangle =
    1 -
    2
    \left\langle \left\| \mathbf{B} - \frac{1}{\tilde B^2} \mathbf{\tilde B}(\mathbf{B} \cdot \mathbf{\tilde B}) \right\|^4 \right\rangle
    \tau^4 + O(B^6 \tau^6).
\]
The extant moments of $\mathbf{B}$ are then replaced as in the case of perpendicular pulses.

\subsection{Pointing instability}
\label{delta}

The process fidelity of an $X$ gate affected by a pointing instability $\delta$, when we estimate this parameter to be $\bar \delta$, is given by the exact expression:
\begin{align*}
    F = &
    \left|
      \mathrm{Tr} \left\{
      \exp \left\{
        - i \frac{\pi}{2} \left(
            \frac{1-\delta^2}{1-\bar\delta^2} - 1
          \right) \sigma_x
      \right\}
      \right\}
    \right|^2 \\
    = & \frac{1}{2}
      + \frac{1}{2} \cos \left(
      \pi
      \frac{\bar\delta^2-\delta^2}{1-\bar\delta^2}
    \right).
\end{align*}
Multiplying by the probability distribution and integrating, we find the following expression for the average process fidelity:
\begin{multline*}
    \langle F_\text{nospec} \rangle
    = \frac{1}{2}
    + \frac{1}{2}
    \sqrt{ \frac{1-\bar\delta^2}{1-\bar\delta^2+2i N(\Delta \delta)^2 \pi} }
    \\ \times
    \mathrm{Re} \left\{
    \exp \left\{
        \frac{-i \delta_0^2 \pi}{1-\bar\delta^2+2i N(\Delta \delta)^2 \pi}
        + \frac{i \bar \delta^2 \pi}{1-\bar\delta^2}
    \right\} \right\}.
\end{multline*}
and where $\delta_0$ is the initial value, and the $\Delta \delta$ is the standard deviation of the Gaussian steps of the random walk.

\subsection{Amplitude instability}
\label{varepsilon}

If we apply an X gate affected by the $\varepsilon$ parameter via an SK1 pulse sequence, the process fidelity will be:
\begin{multline*}
    F =
      \left[ \frac{7}{8} + \frac{1}{8} \cos^2 \left(\pi \frac{\bar \varepsilon - \varepsilon}{1 - \bar \varepsilon} \right) \right]^2
        \cos^2 \left( \frac{\pi}{2} \frac{\bar \varepsilon - \varepsilon}{1 - \bar \varepsilon} \right)
      \\+ \frac{1}{4} \sin\left(2\pi \frac{\bar \varepsilon - \varepsilon}{1 - \bar \varepsilon}\right)
        \sin \left( \pi \frac{\bar \varepsilon - \varepsilon}{1 - \bar \varepsilon} \right)
        \left[ \frac{7}{8} + \frac{1}{8} \cos^2 \left(\pi \frac{\bar \varepsilon - \varepsilon}{1 - \bar \varepsilon} \right) \right]
      \\+ \frac{1}{16} \sin^2 \left(2\pi\frac{\bar \varepsilon - \varepsilon}{1 - \bar \varepsilon}\right)
        \sin^2 \left( \frac{\pi}{2} \frac{\bar \varepsilon - \varepsilon}{1 - \bar \varepsilon} \right),
\end{multline*}
where $\bar\varepsilon$ is the current estimate of the error parameter.
Multiplying by the probability distribution, writing all the trigonometric functions in terms of complex exponentials, and integrating, we find the following final expression:
\[
    \langle F_\text{nospec} \rangle =
    \frac{9}{2^{11}} C_5
    - \frac{15}{2^{10}} C_4
    - \frac{155}{2^{11}} C_3
    + \frac{15}{2^{8}} C_2
    + \frac{585}{2^{10}} C_1
    + \frac{467}{2^{10}},
\]
where the $C_q$ are defined as the real part of a product of exponentials:
\[
    C_q =
    \mathrm{Re} \left\{
    e^{-q^2 N\pi^2 (\Delta \varepsilon)^2/(2(1-\bar\varepsilon)^2)}
    e^{-iq \pi (\bar\varepsilon - \varepsilon_0)/(1-\bar\varepsilon)}
    \right\}.
\]
As usual, $\varepsilon_0$ is the initial value, and the $\Delta \varepsilon$ is the
standard deviation of the Gaussian steps of the random walk.

\section{Analytical expressions for the fidelity with spectator qubits}
\label{app:withspec}

\subsection{Pointing instability}
\label{app:specdelta}

We find an approximate analytical expression for the fidelity after $k$ spectator cycles by integrating the formula for the fidelity given in Supplementary Section III.C according to Eq. (26).
To calculate the integrals, we need the Fisher information, which we find from Eq. (3) with $\mathbf{\hat n} \cdot \boldsymbol\sigma/2$ replaced with the terms that multiply $\delta$ in Eq. (19):
\[
    \frac{\Omega t}{c}  \pm 2 \sqrt{\ln c} =
    \pm \frac{4\pi \sqrt{\ln c}}{c},
\]
where we choose $\Omega t = 4 (\pi/2)$, which is the ideal duration of four $X$ gates in sequence.
We then find the following value for the Fisher information after $M$ measurements of two qubits:
\[
    f_\delta =
    2 M \ln c \left( \frac{8\pi}{c} \frac{1}{1- \bar \delta_{(k-1)M}^2} \right)^2
    \approx
    2 M \ln c \left( \frac{8\pi}{c} \right)^2,
\]
where we are neglecting the effect that the quadratic correction of our previous estimate $\bar \delta_{(k-1)M}$ will have in the Fisher information.
Calling:
\begin{align*}
    J = & \frac{1}{2kM(\Delta \delta)^2}, \\
    K = & \frac{M-1}{2kM}, \\
    W = & 1+2iN\sigma^2 \pi,
\end{align*}
we find the following approximate analytical expression for small $\delta$:
\begin{multline*}
  \langle F_\text{spec} \rangle =
  \frac{1}{2} + E \sqrt{J}
  e^{\alpha}
  e^{- \delta_0^2J}
  \sqrt{\left( J-\gamma \right)^{-1}} \\
  \times
  \exp \left\{
    \frac{\left( J \delta_0 +\beta/2 \right)^2}{ J-\gamma }
  \right\}
  \left\{
    A + B \left[
      \frac{J \delta_0+\beta/2}{ J-\gamma }
    \right] \right.
    \\ \left. + C \left[
      \frac{1}{2(J-\gamma)}
      +  \left( \frac{J\delta_0+\beta/2}{J-\gamma} \right)^2
    \right]
  \right\} + O ( \delta_{kM}^3),
\end{multline*}
where in all integrals we discarded terms of higher order than the third power of the variable being integrated.
The parameters $A$, $B$, $C$, $E$, $\alpha$, $\beta$, and $\gamma$ are:
\begin{align*}
  A = &
    1 -i \frac{N (\Delta \delta)^2 \pi}{W}
    \left\{
    \frac{1}{2z_1}
    + \frac{1}{8(\Delta\delta)^4}
    \frac{1}{z_1^2}
    \left[
        w_1
        + \frac{w_1^2}{2\sigma_k^4} J^2 \delta_0^2
    \right]
  \right\}, \\
  B = &
  -i \frac{N (\Delta \delta)^2 \pi}{W}
  \left[
  \frac{f_\delta^2}{8}
  \frac{1}{z_1^2}
     \frac{w_1^2}{\sigma_k^4} \left( 1 - K \right) K \delta_0
  \right], \\
  C = &
  \frac{1}{2} \left[ \frac{w_2}{w_1} - \frac{z_2}{z_1} \right]
  \left\{
  1 -i \frac{N (\Delta \delta)^2 \pi}{W}
  \left\{
  \frac{1}{2z_1} \right. \right. \\
  & \left. \left. +
  \frac{f_\delta^2}{8}
  \frac{1}{z_1^2}
  \left[
      w_1
      + \frac{w_1^2}{2\sigma_k^4} K^2 \delta_0^2
  \right]
  \right\}
  \right\} \\
    & -i \frac{N (\Delta \delta)^2 \pi}{W}
    \left\{
    - \frac{z_2}{2z_1^2}
    + \frac{f_\delta^2}{8} \frac{1}{z_1^2} \left[
      \frac{w_1^2}{2\sigma_k^4} \left( 1 - K \right)^2 \right. \right. \\
      & \left. \left.
      - 2w_1 \frac{z_2}{z_1}
      \left( 1 + \frac{ w_1K^2 \delta_0^2}{2\sigma_k^4} \right)
            + w_2 \left( 1 + \frac{w_1}{\sigma_k^4} K^2 \delta_0^2 \right)
      \right]
    \right\}, \end{align*}\begin{align*}
  E = &
    \frac{f_\delta^2}{4 \bar \sigma} \sqrt{\frac{w_1}{W z_1}}, \\
  \alpha = &
    \left( \frac{w_1}{4\sigma_k^4}- \frac{1}{2\sigma_k^2} \right)
    K^2 \delta_0^2, \\
  \beta = &
    2 \left( \frac{w_1}{4\sigma_k^4}- \frac{1}{2\sigma_k^2} \right)
    \left( 1 - K \right) K \delta_0, \\
  \gamma = &
    - \frac{i\pi}{Z}
    + \left( \frac{w_1}{4\sigma_k^4}- \frac{1}{2\sigma_k^2} \right)
    \left( 1 - K \right)^2
    + \frac{w_2}{4 \sigma_k^2} K^2 \delta_0^2,
\end{align*}
where $\sigma_k^2$ is the variance of the estimator given in Eq. (26), and we defined the complex numbers $w_1$, $w_2$, $z_1$, and $z_2$ as:
\begin{align*}
  w_1 = &
  \frac{4 \sigma_k^2 z_1}
  {2z_1 + 2z_1 \sigma_k^2 f_\delta^2 - f_\delta^4}, \\
  w_ 2 = &
  \frac{4 \sigma_k^4 f_\delta^4 z_2}
  {(2z_1 + 2z_1 \sigma_k^2 f_\delta^2 - \sigma_k^2 f_\delta^4)^2},
\end{align*}
and:
\[
  z = z_1 + z_2 \delta^2_{kM} =
  \left( \frac{f_\delta^2}{2} - i \pi \right)
  + \left( \frac{i\pi}{W^2} \right) \delta^2_{kM}.
\]

\subsection{Amplitude instability}
For the case of $\varepsilon$ errors, we follow the same procedure, so that the Fisher information for two qubits after $M$ measurements is obtained using Eqs. (3) and (19), and assuming a single $X$ gate ($\Omega t = \pi/2$):
\[
    f_\varepsilon =
    2 M \left( \frac{\pi}{c} \frac{1}{1-\bar \varepsilon_{(k-1)M}} \right)^2 \approx 2 M \left( \frac{\pi}{c} \frac{1}{1-\bar \varepsilon_{kM}} \right)^2.
\]
Here, $\bar \varepsilon_{(k-1)M}$ is the estimate of the error parameter in the previous cycle.
For purposes of calculating the Fisher information, we assume that the estimate is not very different from the value in the current cycle, $\bar \varepsilon_{kM}$.

To integrate the expression derived in Supplementary Section III.C according to Eq. (26), we can notice that all we need are the integrals of the $C_q$.
Defining $C^\prime_q$ as the complex number of which $C_q$ is the real part ($C_q = \mathrm{Re} \{ C_q^\prime \}$), we find a solution of the form:
  \begin{multline*}
  \langle C_q^\prime \rangle_{k>0} =
  \frac{e^{F^\prime} e^{E^{\prime2}/(4D^\prime)}}{\sqrt{2 kM D^\prime (\Delta \varepsilon)^2}}
  \frac{1}{\sqrt{\alpha^\prime}} \\
  \times \left[ 1- \frac{1}{2} \frac{\beta^\prime}{\alpha^\prime}
  \frac{E^\prime}{2D^\prime}
  + \frac{3}{8} \frac{\beta^2}{\alpha^2}
  \left( \frac{1}{2D} + \frac{E^{\prime2}}{4D^{\prime2}} \right) \right] + O (\varepsilon_{kM}^3),
  \end{multline*}
  where, once again, we discard terms in the integrand of higher order than the third power of the variable that we are integrating.
  The parameters $\alpha^\prime$, $\beta^\prime$, $D^\prime$, $E^\prime$, and $F^\prime$ are defined differently from the previous appendix:
    \begin{align*}
    \alpha^\prime = &
    1
    + \left[ f_\varepsilon^{-2}
     - \tilde\sigma^2 \right]
     \left[ 3n (q \pi \sigma)^2
    + 2 i q \pi  \right], \\
    \beta^\prime = &
    - 2 i q \pi\left[ f_\varepsilon^{-2}
     - \tilde\sigma^2 \right], \\
    D^\prime = & \frac{1}{2kM\Delta \varepsilon^2} - \sum_{i=1}^5
    \frac{1}{d_i} \left[ c_i
      + a_i \left( \frac{e_i^2}{d_i^2}
      - \frac{f_i}{d_i} \right)
      - \frac{b_i e_i}{d_i}
    \right], \\
    E^\prime = & \frac{\varepsilon_0}{kM\Delta \varepsilon^2} + \sum_{i=0}^5
    \frac{1}{d_i} \left[
      b_i - \frac{a_i e_i}{d_i}
    \right], \\
    F^\prime = & - \frac{\varepsilon_0^2}{2kM \Delta \varepsilon^2} + \sum_{i=0}^5 \frac{a_i}{d_i},
  \end{align*}
  and the $a_i$, $b_i$, $c_i$, $d_i$, $e_i$, and $f_i$, are:
  \begin{align*}
    a_1 = & - \frac{N}{2} (q \pi \Delta \varepsilon)^2, \\
    b_1 = & i q \pi, \\
    d_1 = & 1, \\
    c_1 = & 0 \\
    e_1 = & 0 \\
    f_1 = & 0,
  \end{align*}

\noindent  and

  \begin{align*}
    a_2 = & \frac{1}{2}
    (f_\varepsilon^{-1})^2 \left[
      N (q \pi \Delta \varepsilon)^2
      + i q \pi
    \right]^2, \\
    b_2 = & (f_\varepsilon^{-1})^2 \left[profiad carreg lafar
      N (q \pi \Delta \varepsilon)^2
       + i q \pi
    \right] (-i q \pi), \\
    c_2 = & \frac{(-i q \pi)^2}{2} (f_\varepsilon^{-1})^2, \\
    d_2 = & 1
    + (f_\varepsilon^{-1})^2 \left[ 3N (q \pi \Delta \varepsilon)^2
    + 2 i q \pi
    \right], \\
    e_2 = & -2 i q \pi (f_\varepsilon^{-1})^2,  \\
    f_2 = &  0,
  \end{align*}

\noindent  and

  \begin{align*}
    a_3 = & \frac{1}{2}
    \left( K \varepsilon_0 \right)^2 \left[
      3 N (q \pi \Delta \varepsilon)^2
      + 2i q \pi
    \right], \\
    b_3 = &
    \left( K \varepsilon_0 \right)
    \left( 1-K \right)
    \left[
      3 N (q \pi \Delta \varepsilon)^2
      + 2i q \pi
    \right] \\ &
    + \frac{1}{2} \left( K \varepsilon_0 \right)^2
    (-2iq\pi), \\
    c_3 = & \frac{1}{2}
    \left( 1-K \right)^2
    \left[
      3 N (q \pi \Delta \varepsilon)^2
      + 2i q \pi
    \right] \\ &
    + \left( K \varepsilon_0 \right)
    \left( 1-K \right)
    (-2iq\pi), \\
    d_3 = & 1
    + \left[ (f_\varepsilon^{-1})^2
      - \sigma_k^2 \right] \left[ 3N (q \pi \Delta \varepsilon)^2
    + 2 i q \pi
    \right],  \\
    e_3 = & -2 i q \pi \left[ (f_\varepsilon^{-1})^2 - \sigma_k^2 \right], \\
    f_3 = & 0,
\end{align*}

\noindent and

 \begin{figure*}[htb!]
    \centering
    \includegraphics[width=.85\columnwidth]{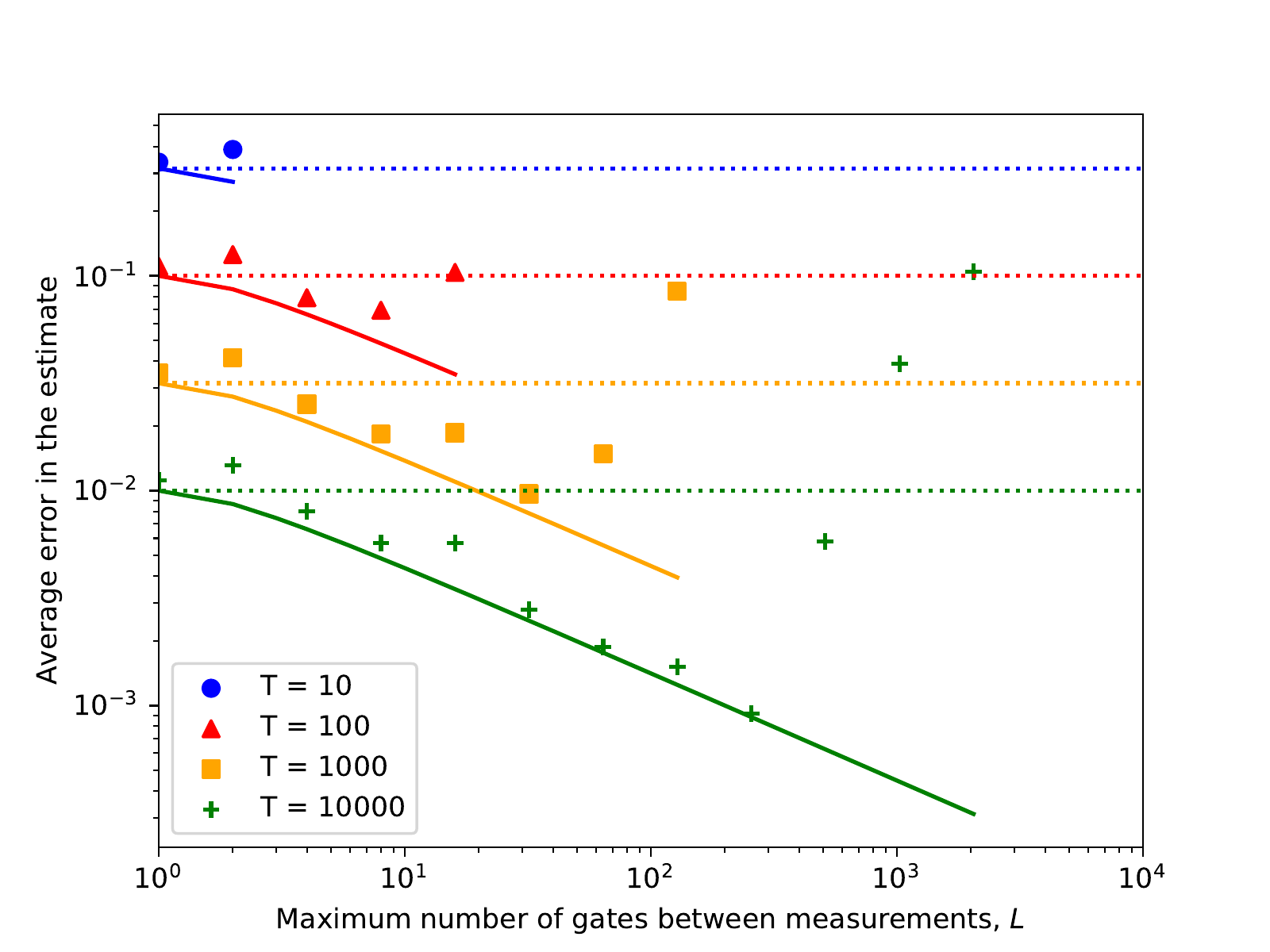}
    \caption{ \textbf{Analysis of RPE under time constraints.}
    If we measure the spectator qubits after each gate during the whole period, our average estimate is lower bounded by the shot-noise limit of $1/\sqrt{T}$, represented by the dashed lines.
    Using RPE, the Cram\' er--Rao bound (solid curves) is saturated as the size of $L$ increases, up to the point when the number of individual experiments becomes too small and the performance becomes as bad or worse than the shot noise limit.
    Numerical simulations for RPE (scattered points) were calculated from an average over 10000 runs.
    }
    \label{fig:rpe}
\end{figure*}

  \begin{align*}
    a_4 = & - \frac{1}{2} \left(K \varepsilon_0 \right)
    \left[
      N (q \pi \Delta \varepsilon)^2
      + i q \pi
    \right], \\
    b_4 = & - \left( 1-K \right) \left[
      N (q \pi \Delta \varepsilon)^2
      + i q \pi
    \right] - \left( K \varepsilon_0 \right)
    (-i q \pi), \\
    c_4 = & - \left( 1-K \right)
    (-i q \pi),\\
    d_4 = & 1
    + \left[ (f_\varepsilon^{-1})^2
      - \sigma_k^2 \right] \left[ 3N (q \pi \Delta \varepsilon)^2
    + 2 i q \pi
    \right], \\
    e_4 = & -2 i q \pi \left[ (f_\varepsilon^{-1})^2 - \sigma_k^2 \right], \\
    f_4 = & 0,
  \end{align*}

  \noindent and

  \begin{align*}
    a_5 = & \frac{\sigma_k^2}{2} \left[
      N (q \pi \Delta \varepsilon)^2
      + i q \pi
    \right]^2, \\
    b_5 = & \sigma_k^2 \left[
      N (q \pi \Delta \varepsilon)^2
      + i q \pi
    \right] (-i q \pi), \\
    c_5 = & \frac{\sigma_k^2}{2} (-i q \pi)^2, \\
    d_5 = & \left[ 1
    + \left[ (f_\varepsilon^{-1})^2
      - \sigma_k^2 \right] \left[ 3N (q \pi \Delta \varepsilon)^2
    + 2 i q \pi
    \right] \right] \\ & \times
    \left[ 1
    + \left[ (f_\varepsilon^{-1})^2 \right] \left[ 3N (q \pi \Delta \varepsilon)^2
    + 2 i q \pi
    \right] \right], \\
    e_5 = &  \left[ 1
    + \left[ (f_\varepsilon^{-1})^2
      - \sigma_k^2 \right] \left[ 3N (q \pi \Delta \varepsilon)^2 \right. \right. \\ &
    \left. \left. + 2 i q \pi
    \right] \right]
    (f_\varepsilon^{-1})^2  (-2 i q \pi) \\
    & + \left[ 1
    + (f_\varepsilon^{-1})^2
      \left[ 3N (q \pi \Delta \varepsilon)^2 \right. \right. \\ &  \left. \left.
    + 2 i q \pi
    \right] \right]
    \left[ (f_\varepsilon^{-1})^2
     - \sigma_k^2 \right] (-2 i q \pi), \\
    f_5 = & (f_\varepsilon^{-1})^2 \left[ (f_\varepsilon^{-1})^2
     - \sigma_k^2 \right] (-2 i q \pi)^2.
  \end{align*}
  In these expressions, the parameter $K$ is defined as in Supplementary Section IV.A, and $\sigma_k^2$ is given in Eq. (26).

{

\section{Performance of RPE under time constraints}
Here we present results for the average imprecision in the estimate of an over-rotation via RPE during a limited time slot in which at most $T$ gates can be applied -- after this period the angle is assumed to change and a new estimate will be required.
The angle of the overrotation is chosen from a normal distribution centered at zero and with $\pi/32$ standard deviation, and happens during a $\pi/2$ $X$ rotation.

In the situation where the time is limited because the overrotation parameters themselves are changing, we see in Fig. \ref{fig:rpe} that increasing the size of the concatenations in RPE improves the precision in the limit of the Cram\' er--Rao bound up until the point where the number of measurements becomes too small.
After that, the scheme breaks down, and the precision of the estimate becomes as bad or worse than the approach without RPE.
Although Fig. \ref{fig:rpe} shows that RPE has potential to improve the precision of estimates for overrotations that change in time, it also shows that its benefits are not as clear as the case where the values are static.

}